\documentstyle[sprocl]{article}

\input{psfig}
\def\be{\begin{equation}}
\def\ee{\end{equation}}
\def\bea{\begin{eqnarray}}
\def\eea{\end{eqnarray}}

\newcounter{dafigcounter}
\def\thedafigcounter{\arabic{dafigcounter}}
\newcommand{\widefig}[3]{\refstepcounter{dafigcounter}
\begin{minipage}{\textwidth}
\begin{center}\mbox{#1}\end{center}
\label{#3}\small \bf Figure \thedafigcounter :\rm\ #2
\end{minipage}\vskip 5mm}

\newcommand{\swfig}[4]{\refstepcounter{dafigcounter}
\begin{minipage}[b]{#2}
\begin{center}
\mbox{#1}
\end{center}
\label{#3}\small \bf Figure \thedafigcounter :\rm\ #4
\end{minipage}}

\begin{document}

\title{HYPERMATTER\\ -PROPERTIES AND FORMATION \\ IN RELATIVISTIC NUCLEAR
COLLISIONS}

\author{L. Gerland, C. Spieles, M. Bleicher, P. Papazoglou,
J. Brachmann, A. Dumitru,\\ H. St\"ocker, W. Greiner}

\address{Institut f\"ur Theoretische Physik,
J.W. Goethe-Universit\"at\\ D-60054 Frankfurt, Germany}

\author{J. Schaffner}

\address{Niels Bohr Institute, Blegdamsvej 17\\ DK-2100 Copenhagen, Denmark}

\author{C. Greiner}   

\address{Institut f\"ur Theoretische Physik,
J.~Liebig-Universit\"at,\\
D-35392 Giessen, Germany}


\maketitle\abstracts{
The extension of the Periodic System into hitherto unexplored domains - antimatter
and hypermatter - is discussed. Starting from an analysis of hyperon and single
hypernuclear properties we investigate the structure of multi-hyperon objects (MEMOs)
using an extended relativistic meson field theory. These are contrasted with multi-strange
quark states (strangelets). Their production mechanism is studied for relativistic
collisions of heavy ions from present day experiments at AGS and SPS to future 
opportunities at RHIC and LHC. It is pointed out that absolutely stable hypermatter
is unlikely to be produced in heavy ion collisions. New attention should be focused on short
lived metastable hyperclusters ($\tau \propto 10^{-10}$s) and on intensity interferometry of 
multi-strange-baryon correlations.}

\section{Motivation}

Since Mendelejew's and Meyer's discovery of regular patterns in the periodic system 
of elements (then only 60 elements were known) has the systematic discovery of new elements 
been a vital motivation for science.

\subsection{Extension of the Periodic System and Nuclear Phasediagram}

The synthesis of new elements 107 - 111 using the cold fusion method at GSI during the last decade 
represents  milestones on the way to a possible superheavy island around $Z \approx 114$~\cite{greiner,sigurd}. 
Fig.~\ref{period} shows the chart of nuclei, as of january 1996, with the new
elements at the far right end. Exotic isotopes as e.g.\ ${}^{100}$Sn,
${}^{151}$Ln, ${}^{11}$Li, discovered at GSI, CAEN, LBL and Riken, are also shown. 
The magic numbers for protons and neutrons and the   
proton- and neutron drip lines are indicated. The crosses ($\times$) show
the large number of newly discovered isotopes, extending the chart of nuclei
from the stable valley outwards to the "shallow waters" of the drip lines.\\
\widefig{\psfig{figure=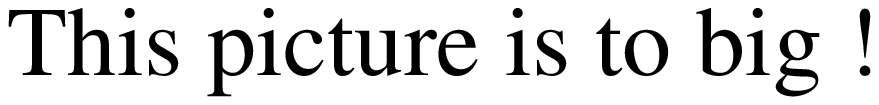,height=3in}}
{The full chart of nuclei}
{period}
Due to the fermi-energy, 
nuclei tend to consist of as many protons as neutrons. However, with increasing charge the nuclei become unstable
due to the Coulomb-interaction: the chart of nuclei is limited in each direction
for nuclei which are "built" of protons and neutrons.\\
 However, here we shall discuss whether
it could be possible to extend the chart of nuclei in new domains, using new degrees of freedom,
namely strangeness!

\subsubsection{Properties of Hyperons}

The flavour  SU(3) - triplet contains u-, d- and s-quarks. Ordinary matter
is solely built of u- and
d-quarks. It is therefore an intriguing thought to search for matter with
s-quarks,
which we shall
call {\bf strange matter}, if it contains sufficiently many s-quarks
per u- and d-quarks.
Fig.~\ref{octet} shows the quark-content of spin 1/2
nucleons and
hyperons contained in the octet and of spin 3/2 baryons contained in the
decouplet. Table 1 reviews properties of these baryons. 
\\
\widefig{\psfig{figure=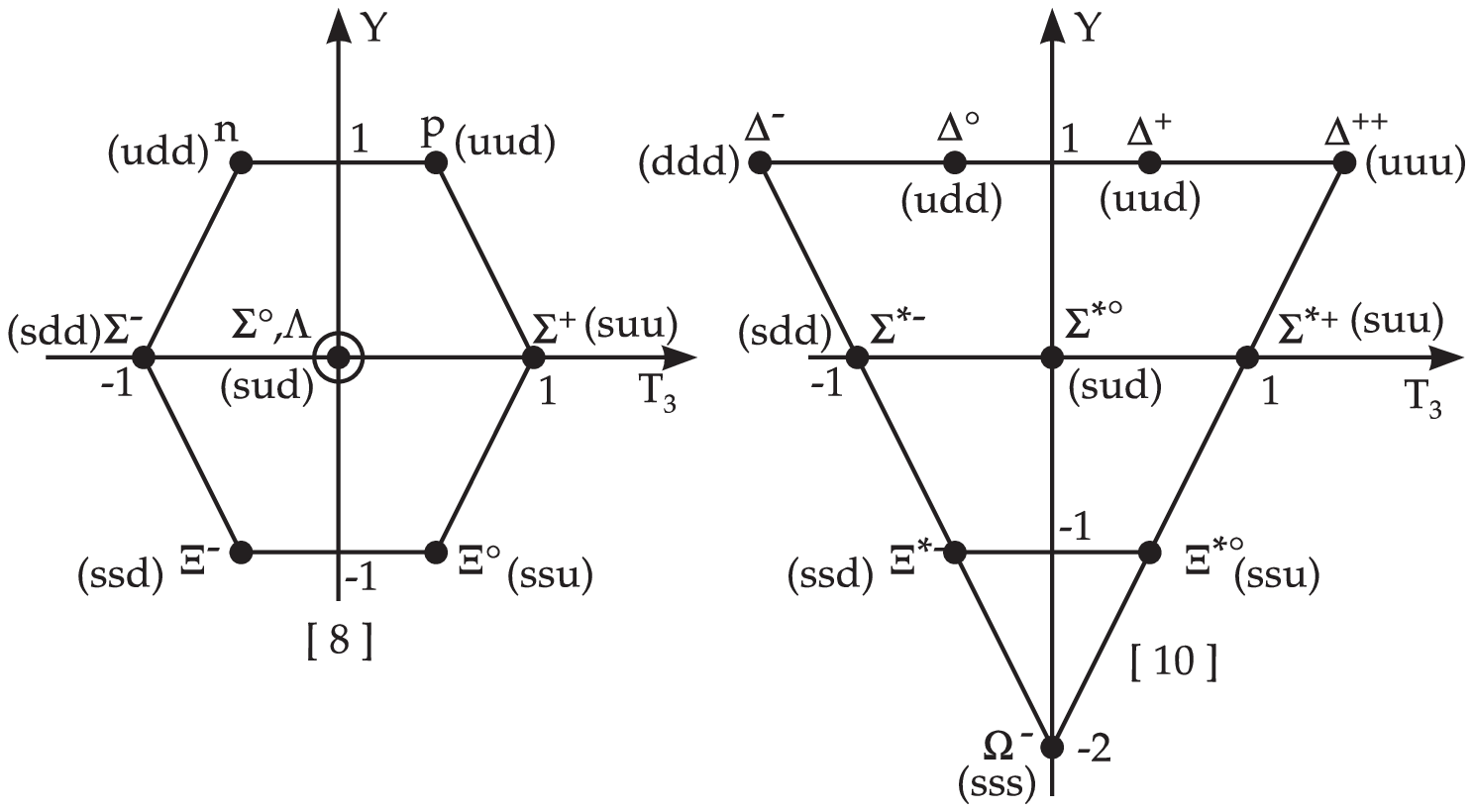,height=2.5in}}%
{The octet of
spin 1/2 nucleons and hyperons (left) and the decuplet of
spin 3/2 baryons (right). The quark content of the particles is indicated.}
{octet}
{\small
\begin{minipage}[t]{11.5cm}
\begin{center}
\begin{tabular}{|c|c|c|c|c|c|l|}
\hline
&J$^{\rm P}$&T&S[C]&mass[MeV]&lifetime[s]&main decays\\
\hline
p&$1/2^+$&1/2&0&938.27&$>10^{25}$years&\quad stable\\
\hline
n&$1/2^+$&1/2&0&939.56&$887.0\pm2.0$&
$  {\rm p}e^- \overline{\nu}_e (100\%)$\\
\hline
$\Lambda$&$1/2^+$&0&-1&1115.7&$2.63\times 10^{-10}$&
$  {\rm p}\pi^-(64.1\%),$\\&&&&&& $ {\rm n}\pi^0(35.7\%)$\\
\hline
$\Sigma^+$&$1/2^+$&1&-1&1189.4&$0.799\times 10^{-10}$&
$  {\rm p}\pi^0  (51.6\%),$\\&&&&&& $ {\rm n}\pi^+  (48.3\%)$\\
\hline
$\Sigma^0$&$1/2^+$&1&-1&1192.6&$7.4\times 10^{-20}$&
$  \Lambda \gamma  (100\%)$\\
\hline
$\Sigma^-$&$1/2^+$&1&-1&1197.4&$1.48\times 10^{-10}$&
$  {\rm n}\pi^-  (99.8\%)$\\
\hline
$\Xi^0$&$1/2^+$&1/2&-2&1314.9&$2.90\times 10^{-10}$&
$  \Lambda\pi^0  (100\%)$\\
\hline
$\Xi^-$&$1/2^+$&1/2&-2&1321.3&$1.64\times 10^{-10}$&
$  \Lambda\pi^-  (100\%)$\\
\hline
$\Omega^-$&$3/2^+$&0&-3&1672.4&$0.82\times 10^{-10}$&
$  \Lambda K^- (67.8\%),$\\
&&&&&& $ \Xi^-\pi^0  (8.6\%),$\\&&&&&& $ \Xi^0\pi^-  (23.6\%)$\\
\hline
$\Lambda_c^+$&$1/2^+$&0&0[1]&2285&$2.0\times 10^{-13}$&\\
\hline
$\Xi_c^+$&$1/2^+$&1/2&-1[1]&2466&$3.5\times 10^{-13}$&\\
\hline
$\Xi_c^0$&$1/2^+$&1/2&-1[1]&2470&$1.0\times 10^{-13}$&\\
\hline
$\Lambda_b^0$&$1/2^+$&0&B=-1&5641&$1.1\times 10^{-12}$ &\\
\hline
\end{tabular}
\end{center}
{\bf Table 1 :} Properties of "stable" baryons and their weak decay modes.
$J$ stands for
spin, $P$ for
parity and $T$ for isospin. The branching ratios of the
decays of particles
with
charm (bottom) are not
yet well known.
\label{tabe}
\cite{padagr}
\vspace{0.2cm}
\end{minipage}}

\subsubsection{Strange Nuclei}

There are various ways to venture from ordinary matter (built out of protons ($Z$)
and neutrons ($N$)
in nuclei with baryon number $A= Z+N$) into
the sector of strange matter(Fig.~\ref{phase}).\\
\swfig{\psfig{figure=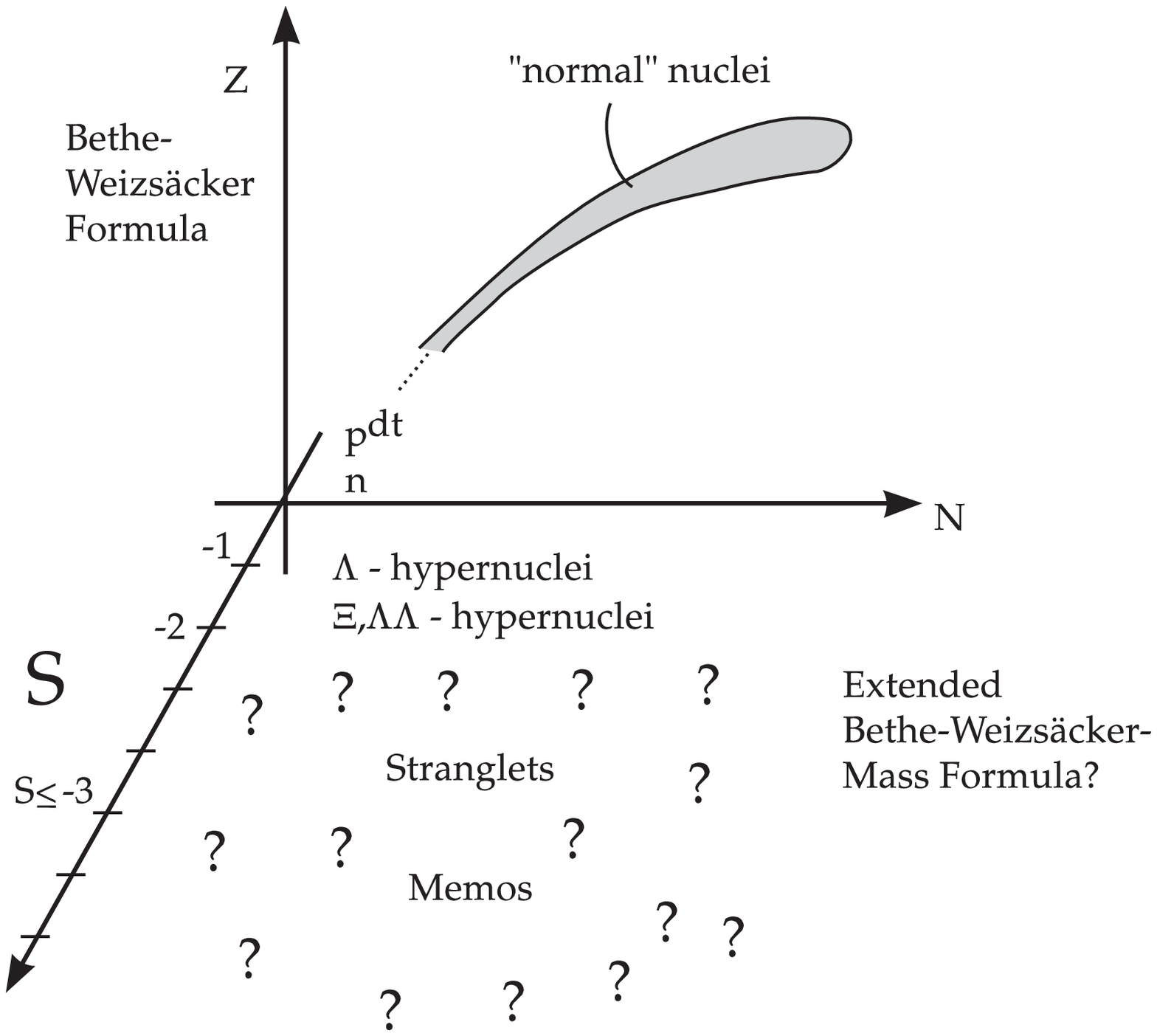,height=2.4in}}
{5.5cm}
{phase}{Schematic view of the novel extension of the
Periodic System (proton-, $Z$,
and neutron number, $N$, plane) into the direction of finite net strangeness $S$}
\hfill
\swfig{\psfig{figure=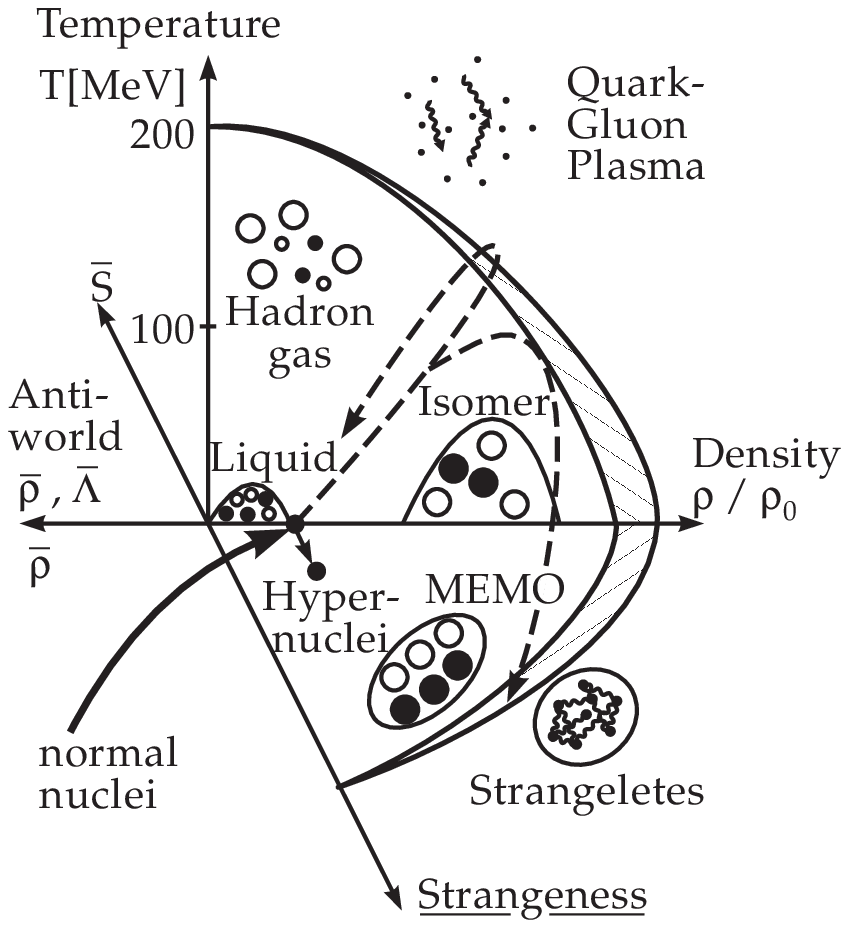,height=2.3in}}
{5.5cm}{phased}{Sketch of the nuclear matter phase diagramm
\newline
\newline 
\newline}\\
One possibility is to add baryons with strangeness
($\Lambda,\,\Xi,\ldots$) to ordinary nuclei.\\
However, strange baryons are short lived, therefore adding consecutively
stran- geness to hypernuclei (e.g. with kaon beams) seems unfeasable. Only in the
 interior of neutronstars or in relativistic heavy ion collisions
one can expect the simultaneous presence and density of sufficiently many hyperons/strange quarks
to allow for a search for multi-strange matter.
 Baryons and mesons in dense matter may also
dissolve and form "{\bf strangelets}" of u-, d-, and s- quarks.

\subsection{Heavy Ion Collisions}

Relativistic heavy ion collisions are the only tool to probe
hot and dense nuclear matter, where one might
create such new forms of nuclear matter, i.e. MEMOs (\underline{M}etastable \underline{E}xotic
\underline{M}ultistrange \underline{O}bject) and strangelets. Fig.~\ref{phased}
shows different paths for nuclear matter in the phase diagram in
heavy ion collisions. Starting at the ground state of ordinary
nuclear matter, at $\rho_0$ and $T=0$, various excursions through that phase diagram
are indicated.
As a result of such collisions ordinary elementary particles of all
kinds are produced. They constitute the "ashes" of what was   
encountered during the high density state of the reactions.\\
Can such matter be stable? This question can be studied theoretically by extending
well tested models of nuclear structure into the strangeness domain.

\section{Stability and Existence of Strange Matter}

Here we study the possibilities for extending the chart of nuclei into these new domains 
by following our knowledge of normal 
nuclei. We use the \underline{R}elativistic \underline{M}ean
\underline{F}ield-(RMF-)model
which has been applied successfully 
in numerous applications to normal nuclei. It consists of an effective Meson-Baryon Lagrangean
with scalar ($\sigma$) and vector ($\omega , \rho$) fields.

\subsection{MEMOs - Strange Nuclei}

\subsubsection{Calculation of Nuclei - RMFT: ${}^{100}$Sn}

An example for the successful application of the RMF-model to exotic nuclear states
is the prediction of 
the shell model potential and level structure for the exotic ${}^{100}$Sn-isotope shown in Fig.~\ref{fig38}.
It reproduces the experimentally measured Q-value of this exotic heaviest
$Z=N$ doubly closed shell nucleus.\\
\\
\widefig{\psfig{figure=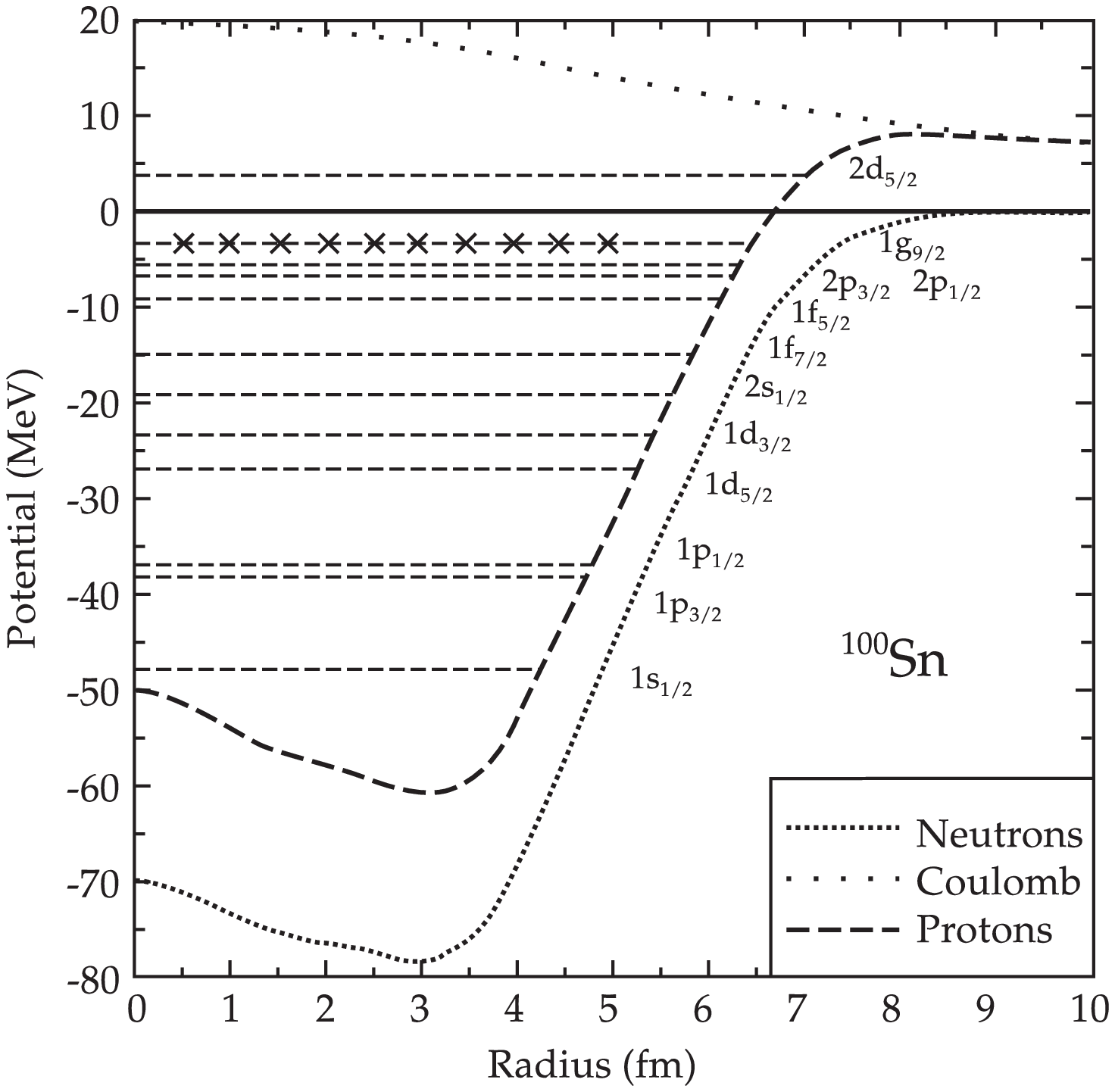,height=2.5in}}%
{Prediction of the shell model potential and level structure for the
exotic ${}^{100}$Sn-isotope. Proton- and neutron potentials are shown~\cite{Sch94}.}
{fig38}

\subsubsection{Extension of RMFT to Hyperons}

This RMF-model has been extended to include single $\Lambda$-hypernuclei~\cite{boguta}.
Fig.~\ref{hypdat} shows the measured single particle energy for
different single $\Lambda$-hypernuclei in comparison with RMF calculations
using only two free parameters,   
namely the two coupling constants of the $\Lambda$
($g_{\sigma\Lambda}$ and $g_{\omega\Lambda}$), which are correlated:
they are fixed by the potential depth of the $\Lambda$
\be
U_\Lambda = - g_{\sigma\Lambda}\sigma - g_{\omega\Lambda} V_0
\ee
in saturated nuclear matter \cite{Sch92}.\\
\\
\widefig{\psfig{figure=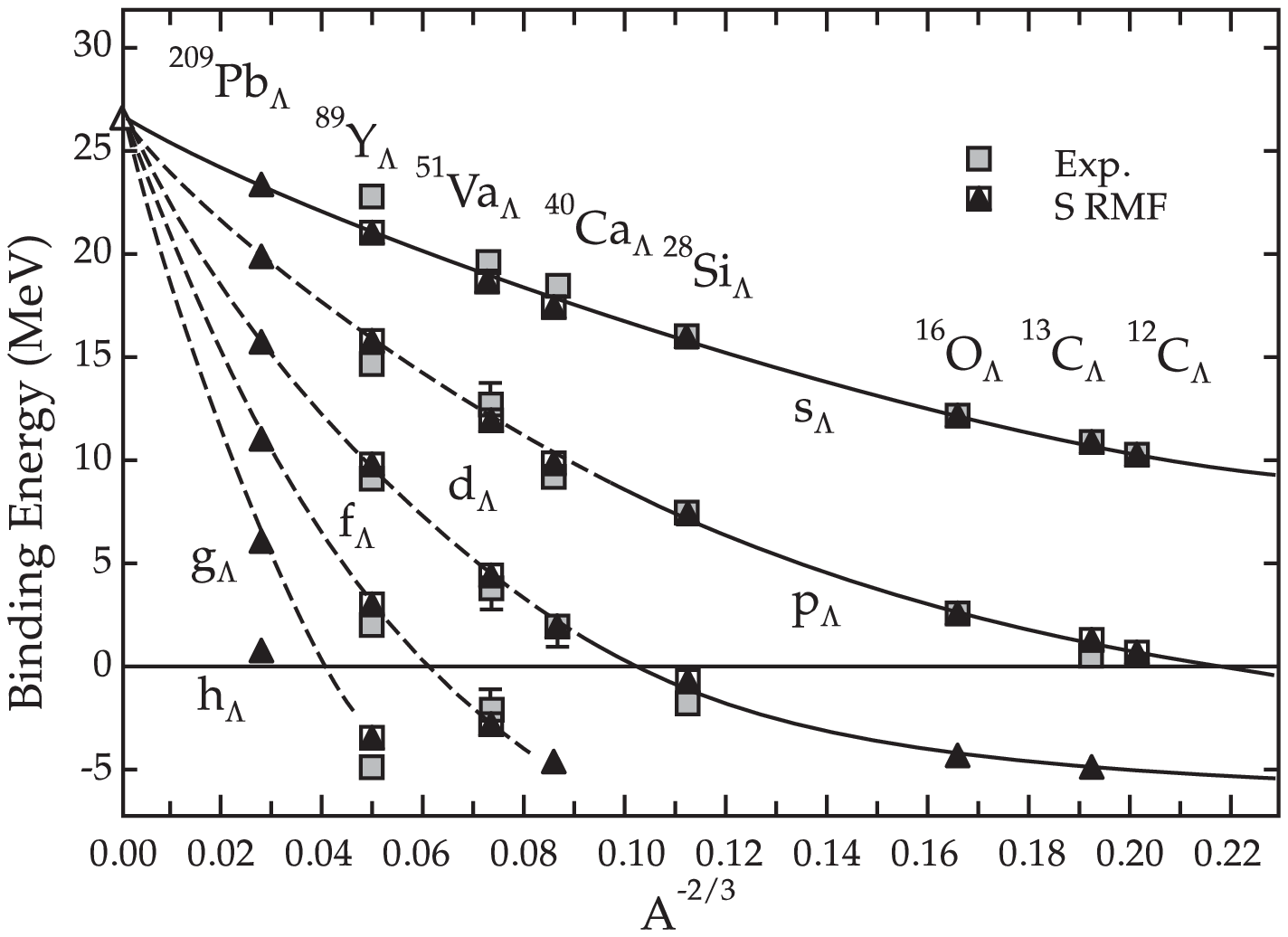,height=2.5in}}%
{The single particle energy of several hypernuclei~\cite{Sch93} in
comparison with RMF calculations.}
{hypdat}
 One can choose for all spin 1/2 baryons,
for example, SU(6)-symmetry for the (u,d)-quark content dependent vector coupling constants to the $\omega$-field
\be
g_{\omega N} = \frac{3}{2} g_{\omega\Lambda}
= \frac{3}{2} g_{\omega\Sigma} =  3g_{\omega\Xi}.
\ee
The scalar coupling constants can be fitted to the potential depth of the
corresponding hyperon.

A $\chi^2-$fit of the coupling constants to experimental data on single-$\Lambda$-hypernuclei is shown in Fig.~\ref{cc}.
Note that only {\bf ratios} of $g_{\sigma\Lambda}$ and $g_{\omega\Lambda}$ can be fixed, but there is a
considerable uncertainty in fixing just one parameter.
\\
\widefig{\psfig{figure=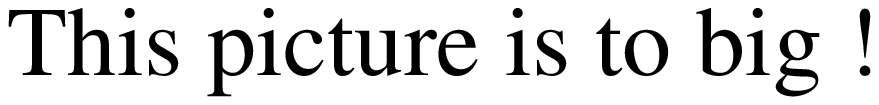,height=2.5in}}%
{$\chi^2$-Fit of the coupling constants to experimental data on single-$\Lambda$-hypernuclei}
{cc}

Typical potentials for $\Lambda$'s and nucleons are given in Fig.~\ref{fig43}.
Obviously the spin-orbit-splitting is
much smaller for $\Lambda$'s as compared to that of the nucleons. Also the potential
is $U_{\Lambda}\approx 1/\!2 U_N$.
\\
\widefig{\psfig{figure=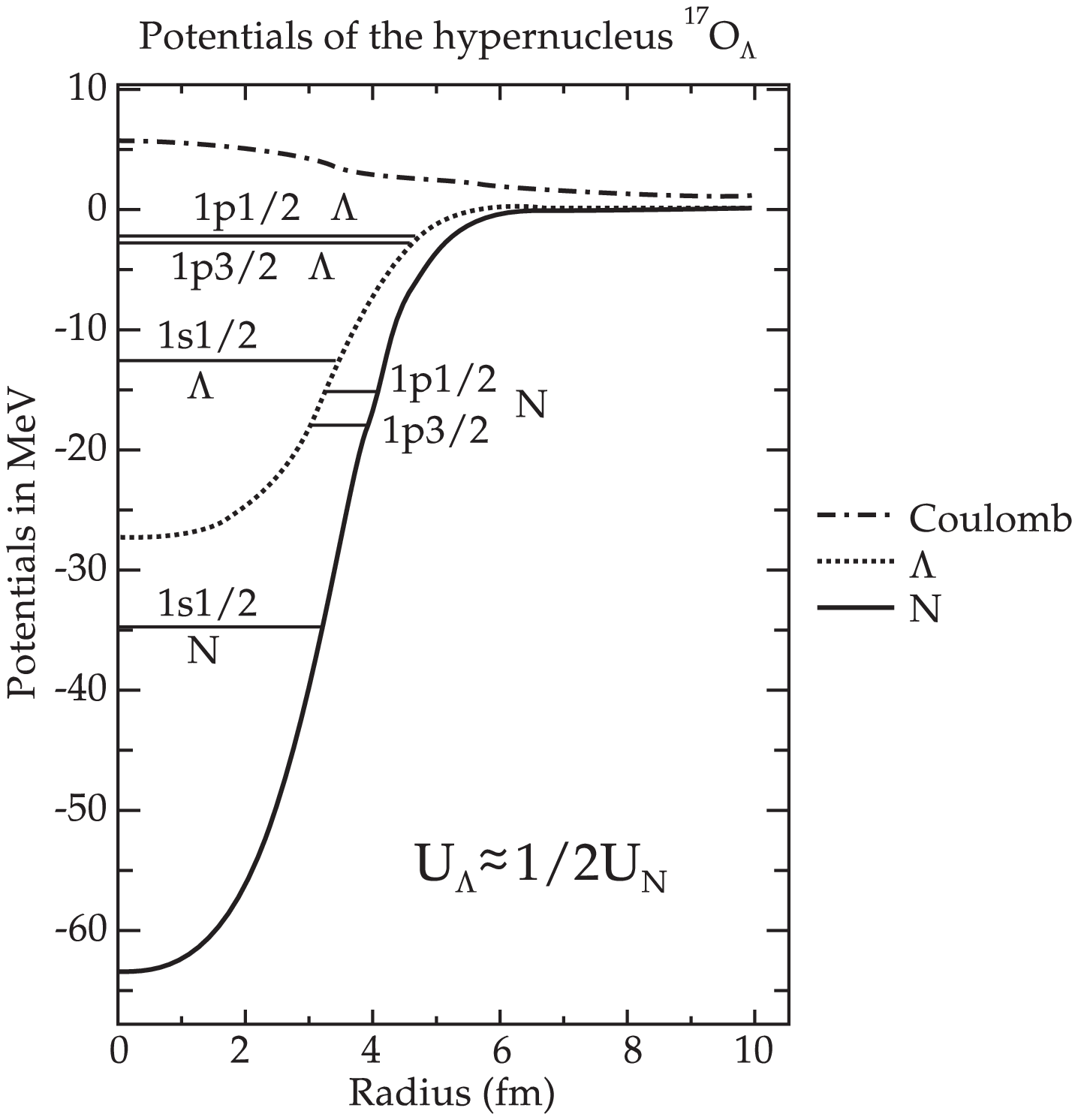,height=2.5in}}%
{Typical potential wells for the $\Lambda$s and for nucleons in case
of the hypernucleus ${}^{17}$O. Note that approximately $U_{\Lambda}\approx 1/2U_N$.}
{fig43}

This model 1 reproduces well the properties of single $\Lambda$-hypernuclei. However, 
as we willdiscuss below, with this model it is not able
to reproduce the observed strongly attractive $\Lambda\Lambda$ interaction.
(The situation can be remedied~\cite{Sch93,Sch94}
by introducing two additional meson fields, the scalar meson $f_0(975)$
(denoted as $\sigma^*$ in the following) and the vector meson
$\phi(1020)$(model 2).)

\subsubsection{Extension to Light Hypernuclei}

Now that the coupling constants are determined from experiment,
the RMF-model can be tested by $ predicting $ the properties of light (A$<16$)
single $\Lambda$ hypernuclei. This serves as a test of the predictive power of the model.
Fig.~\ref{fig42} shows the binding energies of such light hypernuclei as 
measured and as calculated without adjusting any free parameters for the
non-spherical nuclei.
The prediction are within 10\% of the experimental values. This gives confidence
 into using the model in hitherto unexplored domains of hyperon
combinations.\\
\\
\widefig{\psfig{figure=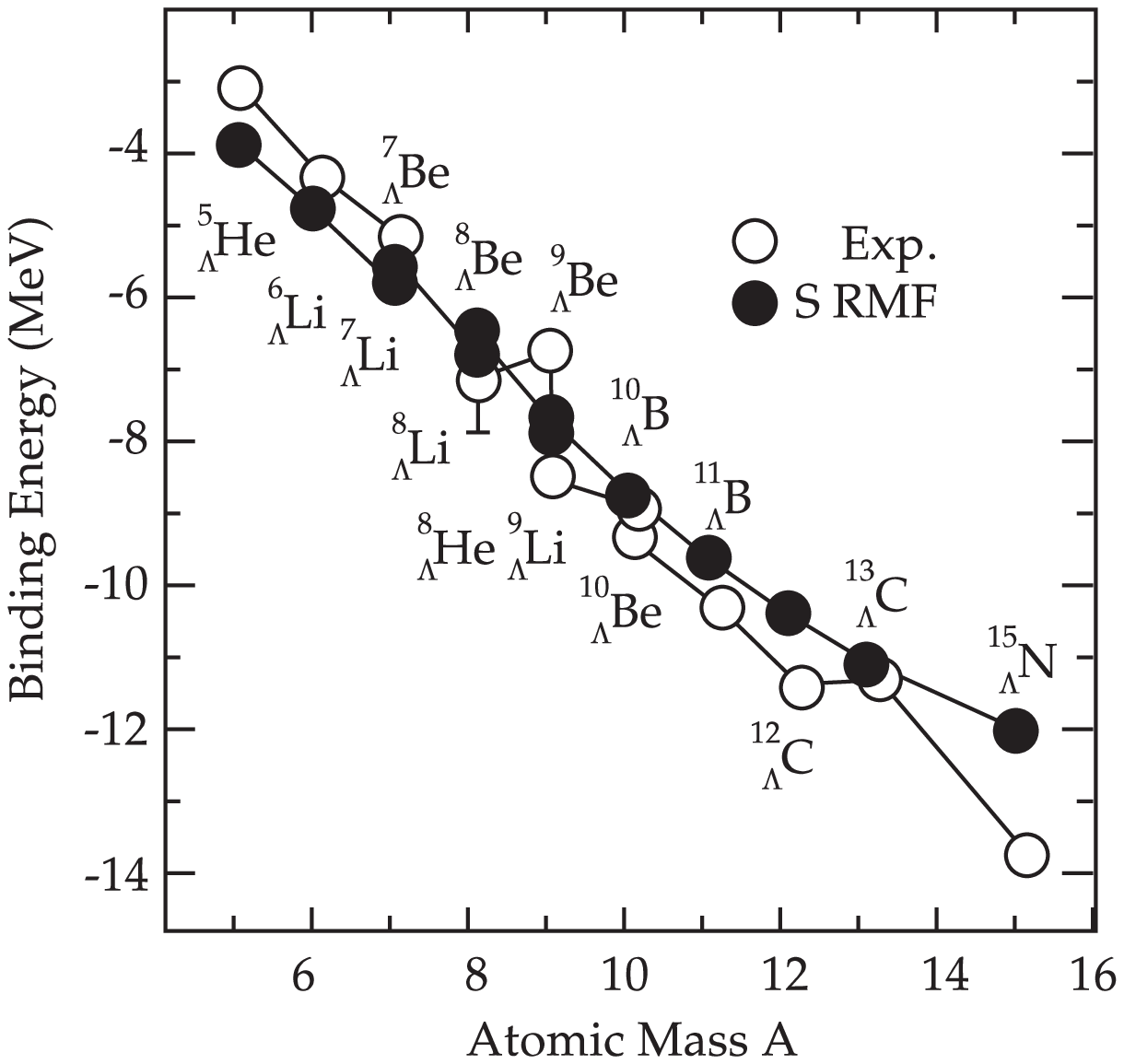,height=2.5in}}%
{Calculated binding energies for a $\Lambda$-particle in a light
nucleus. The experiment is shown in open circles ($\circ$), the theory in
full circles ($\bullet$). The data are from~\cite{crien}.}
{fig42}

\subsubsection{$\Lambda\Lambda$-Interaction}

Such an unexplored domain is the study of multihypernuclei.
Few data are available for double $\Lambda$-nuclei
(for a review see~\cite{Dal89}). Fig.~\ref{pic4} shows that there is a
substantial additional $\Lambda\Lambda$ attraction, not taken care of in the
standard RMF-model. This has been implemented (see model 2 below).
Fig.~\ref{fig44} shows the  density distribution of the double-$\Lambda$
hypernucleus ${}^6{\rm He}_{\Lambda \Lambda}$. Every baryon sits in its respective 
$1s_{1/2}$-state. The $\Lambda$-density is indicated by the dashed area.
Evidently an enhanced interaction radius results. The lifetime of the
$\Lambda$'s is of the order $\tau \approx 10^{-10}$s. This behaviour is that of like the
Z=2 anomalon.\\
\\
\widefig{\psfig{figure=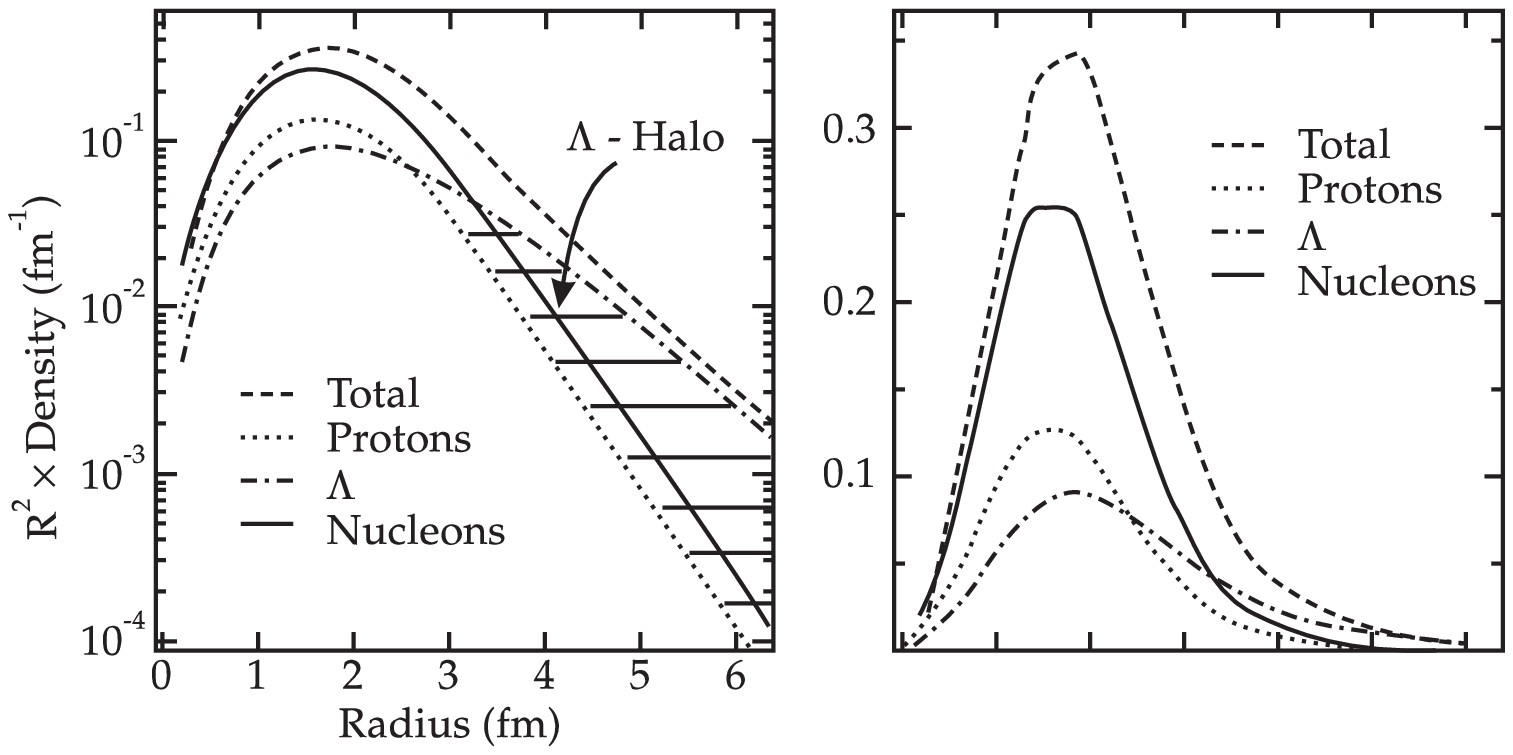,height=2.2in}}%
{ Density distribution of the highly symmetric double-$\Lambda$
hypernucleus ${}^6{\rm He}_{\Lambda \Lambda}$.}
{fig44}

However, it turns out that this model underpredicts the measured 
$\Lambda\Lambda$-interaction for a $\Lambda$-vector-coupling-constant
between 0 and $g_{vN}$ [Fig.~\ref{pic4}]. Therefore~\cite{Sch92}
additional $\Lambda \Lambda$ binding must be implemented into the theory.
In the Mean Field Approximation this can be done e.g. by adding scalar
and vector mesons to the Lagrangean.

The additional Lagrangean then reads
\be
{\cal L}' = {\cal L} + {\cal L}^{YY}_{\rm Meson}
  + {\cal L}^{YY}_{\rm Coupling},
\ee
with
\bea
{\cal L}^{YY}_{\rm Meson} & = &
   \frac{1}{2}\left(\partial_\nu\sigma^*\partial^\nu\sigma^*
     - m_{\sigma^*}^2{\sigma^*}^2\right)
 - \frac{1}{4} S_{\mu\nu}S^{\mu\nu} + \frac{1}{2} m_\phi^2\phi_\mu\phi^\mu,
\\ \cr
{\cal L}^{YY}_{\rm Coupling} & = &
 - \sum_B g_{\sigma^* B}{\overline \Psi}_B\Psi_B\sigma^*
 - \sum_B g_{\phi B}{\overline\Psi}_B\gamma_\mu\Psi_B \phi^\mu.
\eea
The equations of motion for the
additional meson fields
read :
\begin{equation}  
\begin{array}{ccl}
\left(-\Delta+m_{\sigma^*}^2\right)\sigma^*&=&
-g_{\sigma^* \Lambda}\rho_{s\Lambda}
-g_{\sigma^* \Sigma}\rho_{s\Sigma}
-g_{\sigma^* \Xi}\rho_{s\Xi},
 \\
\left(-\Delta+m_\phi ^2\right)\phi_0&=&
+g_{\phi \Lambda}\rho_{v\Lambda}
+g_{\phi \Sigma}\rho_{v\Sigma}
+g_{\phi \Xi}\rho_{v\Xi},
\end{array}   
\end{equation}
with the respective scalar and vector densities. Together with the equations for the $\sigma$-,
$V_0$- and $R_{0,0}$-meson fields and
for the Coulomb field $A_0$, all of which are unchanged,
coupled
system of nonlinear partial differential equations results, which is solved selfconsistently.
The coupling constants are fixed by the $SU(6)$-relations:
\begin{equation}
\frac{g_{\phi\Lambda}}{g_{vN}} = -\frac{\sqrt{2}}{3} \quad , \qquad
\frac{g_{\phi\Xi}}{g_{vN}} = -\frac{2\sqrt{2}}{3}.
\end{equation}

The hyperon-hyperon-interaction is fixed to extrapolated depths of the $U_Y^{Y'}$
-potentials~\cite{Sch94}.

This extended model 2 is now closer to the experimental value of
the $\Lambda\Lambda$-interaction matrix element (see Fig.~\ref{pic4}).\\
\widefig{\psfig{figure=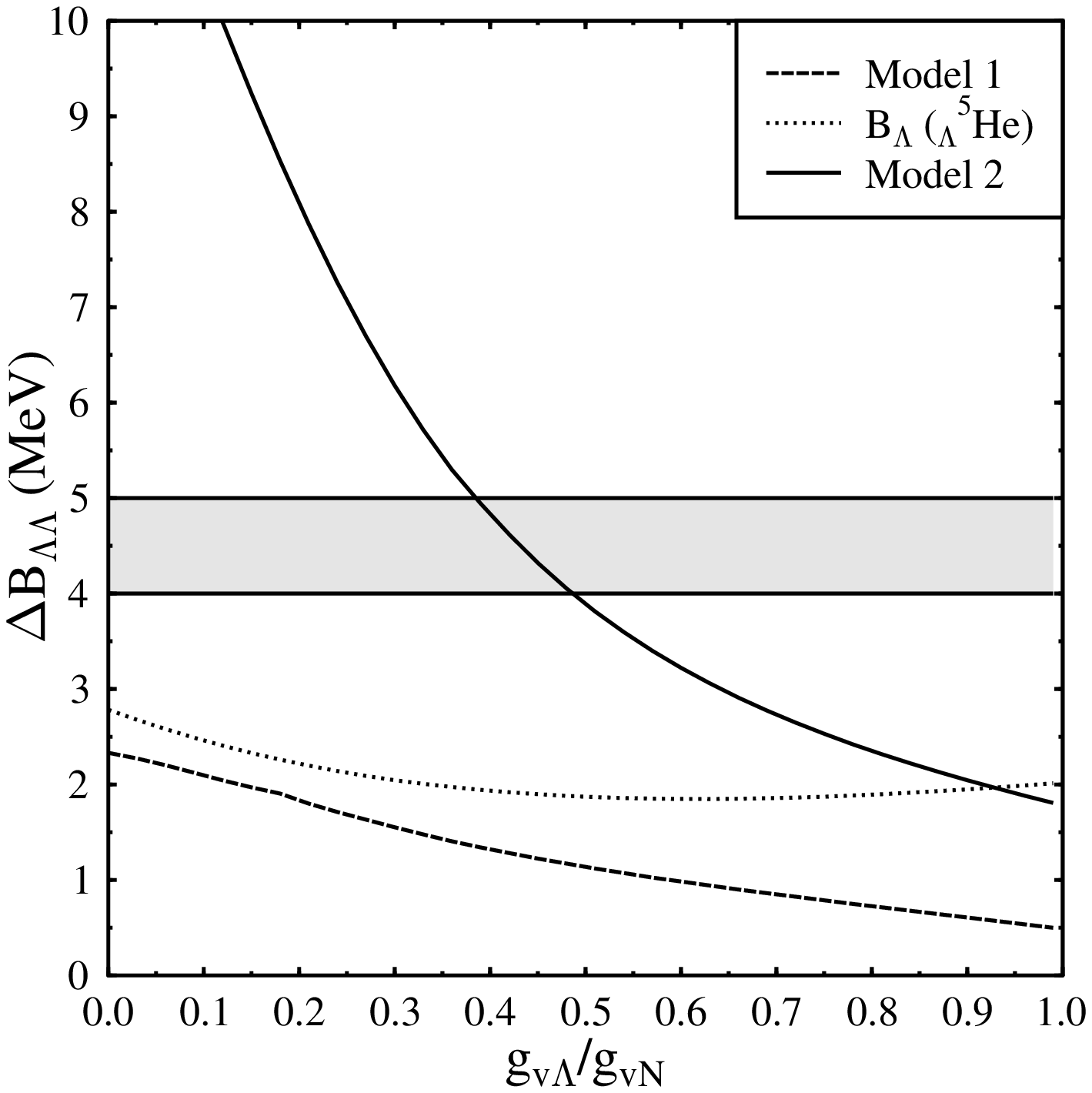,height=2.5in}}%
{The matrix element $\Delta$B$_{\Lambda\Lambda}$ as a function
of the vector coupling constant of the $\Lambda$ in model 1 and 2.}
{pic4}

\subsubsection{Hypermatter -- Stable Combinations of several Hyperons}

Now that all hyper-data known are described, one can go
about extending our chart of nuclei into areas 
{\bf without any} protons and neutrons!
However, keep in mind that strong reactions might change the hyperon-nucleon
composition with millibarn crosssections.

All possible metastable combinations of two baryons are listed in table 2 .\\
\\
\begin{minipage}[t]{11.5cm}
\begin{center}
\begin{tabular}{|c|c|c|c|c|c|}
\hline
$-S\backslash Z$& -2 & -1 & 0 & +1 & +2 \\
\hline
0& & &nn&np&pp\\
\hline
1& &$\Sigma^-$n&$\Lambda$ n&$\Lambda$ p&$\Sigma^+$p\\
\hline
2&$\Sigma^-\Sigma^-$&$\Xi^-$n&$\Lambda\Lambda$&$\Xi^0$p&$\Sigma^+\Sigma^+$\\
\hline
3&$\Xi^-\Sigma^-$&$\Xi^-\Lambda$&$\Xi^0\Lambda$&$\Xi^0\Sigma^+$& \\
\hline
4&$\Xi^-\Xi^-$&$\Xi^0\Xi^-$&$\Xi^0\Xi^0$& & \\
\hline
5&$\Xi^-\Omega^-$&$\Xi^0\Omega^-$& & & \\
\hline
6&$\Omega^-\Omega^-$& & & & \\
\hline
\end{tabular}
\label{tab:dupletts}
\end{center}
{\small {\bf Table 2 :} Possible metastable doublets of the members of the baryon
octet and of the $\Omega^-$.}
\vspace{0.6cm}
\end{minipage}

Metastable (i.e. stable with respect to strong interaction) combinations of three different
species are listed in table 3 .
Of special interest are metastable systems consisting of
$\{ \Lambda,\Xi^0,\Xi^- \}$-hyperons: do they form clusters of purely
hyperonic matter?
More than three different baryon species cannot form a metastable
combination in free space, but binding effects can change this statement in clusters
of hypermatter.\\
\\
\begin{minipage}[t]{11.5cm}
\begin{center}
\begin{tabular}{|c|c|c|c|c|c|}
\hline
$-S\backslash Z$& -2 & -1 & 0 & +1 & +2 \\
\hline
1&&&$\qquad \qquad$ &$\Lambda$np&\\
\hline
2&&&&&\\
\hline  
3&$\Xi^-\Sigma^-$n&$\Xi^-\Lambda$n& &$\Xi^0\Lambda$ p&$\Xi^0\Sigma^+$p\\
\hline
4&&&&&\\
\hline  
5&&$\Xi^-\Xi^0\Lambda$& & &\\
\hline
6& & & & & \\
\hline
7&$\Omega^- \Xi^- \Xi^0$& & & & \\
\hline
\end{tabular}
\end{center} 
{\small 
{\bf Table 3 :} Metastable triplets
of nucleons and hyperons sorted
according to strangeness
$S$ and charge $Z$.
}
\label{tab:tripletts}
\vspace{0.4cm}
\end{minipage}

\subsubsection{MEMOs - Stability due to Pauli Blocking}

Let us turn to the structure of bound exotic multi-strange nuclei. They can be
formed by adding $\Sigma$'s (see \cite{Sch92,Sch93}) and $\Xi$'s. The latter may 
react strongly according to

\begin{equation}
\label{hanswurst}
\begin{array}{lcll}
{\rm n}+\Xi^0 & \rightarrow & 2\Lambda & (\Delta E=23{\rm MeV}) \\
{\rm p}+\Xi^- & \rightarrow & 2\Lambda & (\Delta E=28{\rm MeV}) \\
\end{array}
\end{equation}
However, if there are already several $\Lambda$'s within the nucleus, these
reactions can eventually be
Pauli-blocked. The idea is illustrated in the Fig.~\ref{pauli} for the
MEMO 
$[2{\rm p}\, 2{\rm n}\, 2\Lambda\, 2\Xi^0 2\Xi^-]$, which has
strangeness content $S/A=1$, charge $Z=0$ and density $\rho\approx 4\rho_0$.
The binding
energy difference cancels the mass difference of the strong reaction
channels so that the whole 'hypercluster' is metastable.\\
\widefig{\psfig{figure=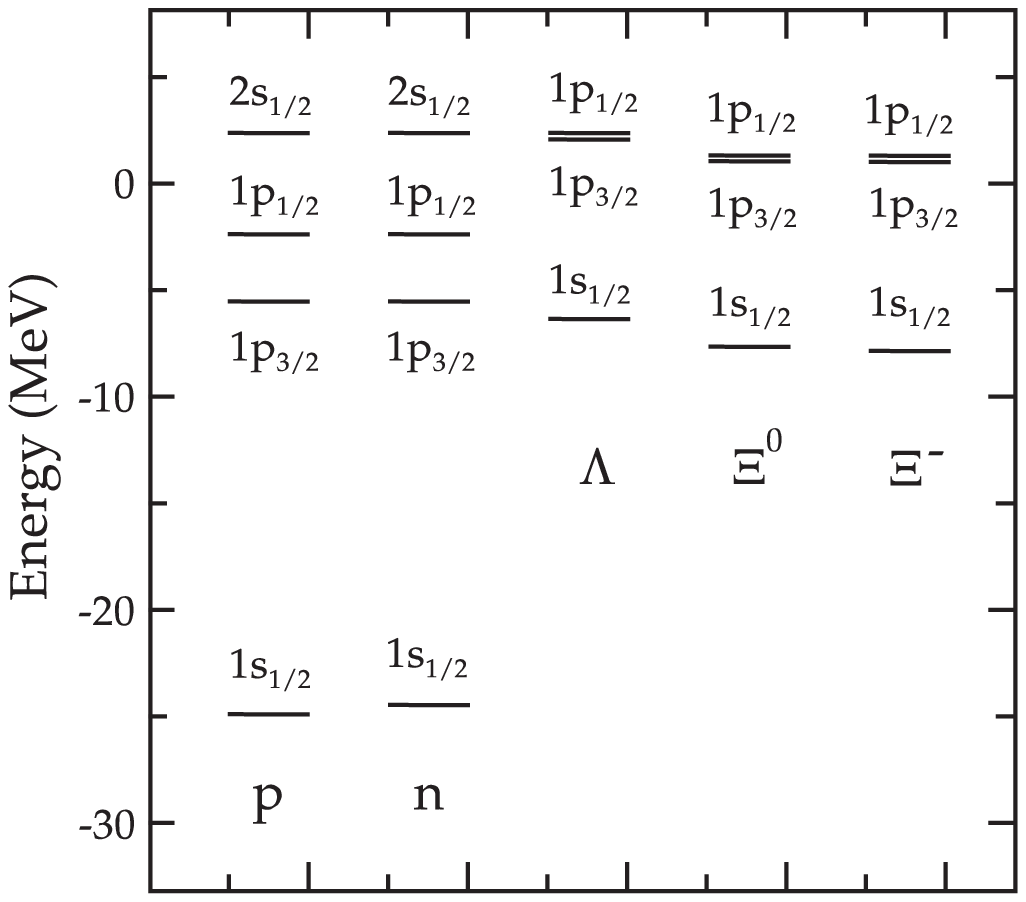,height=2in}}%
{Single particle energy of a MEMO consisting of two of
each baryon of the baryon octet except the $\Sigma$'s.~\cite{Sch93}}
{pauli}
Fig.~\ref{pic5} shows the Q-value of the reaction (\ref{hanswurst}) for the MEMO
${}^7_{\Lambda \Lambda \Xi^0}$He and ${}^7_{\Lambda \Lambda \Xi^{-}}$H. Reaction (\ref{hanswurst}) is forbidden
for negative $\Delta$E values, hence the system can only decay weakly.
The hatched area in Fig.~\ref{pic5} indicates the potential depth for $\Xi$'s as extracted
from $\Xi^{-}$ hypernuclear data.
\\
\widefig{\psfig{figure=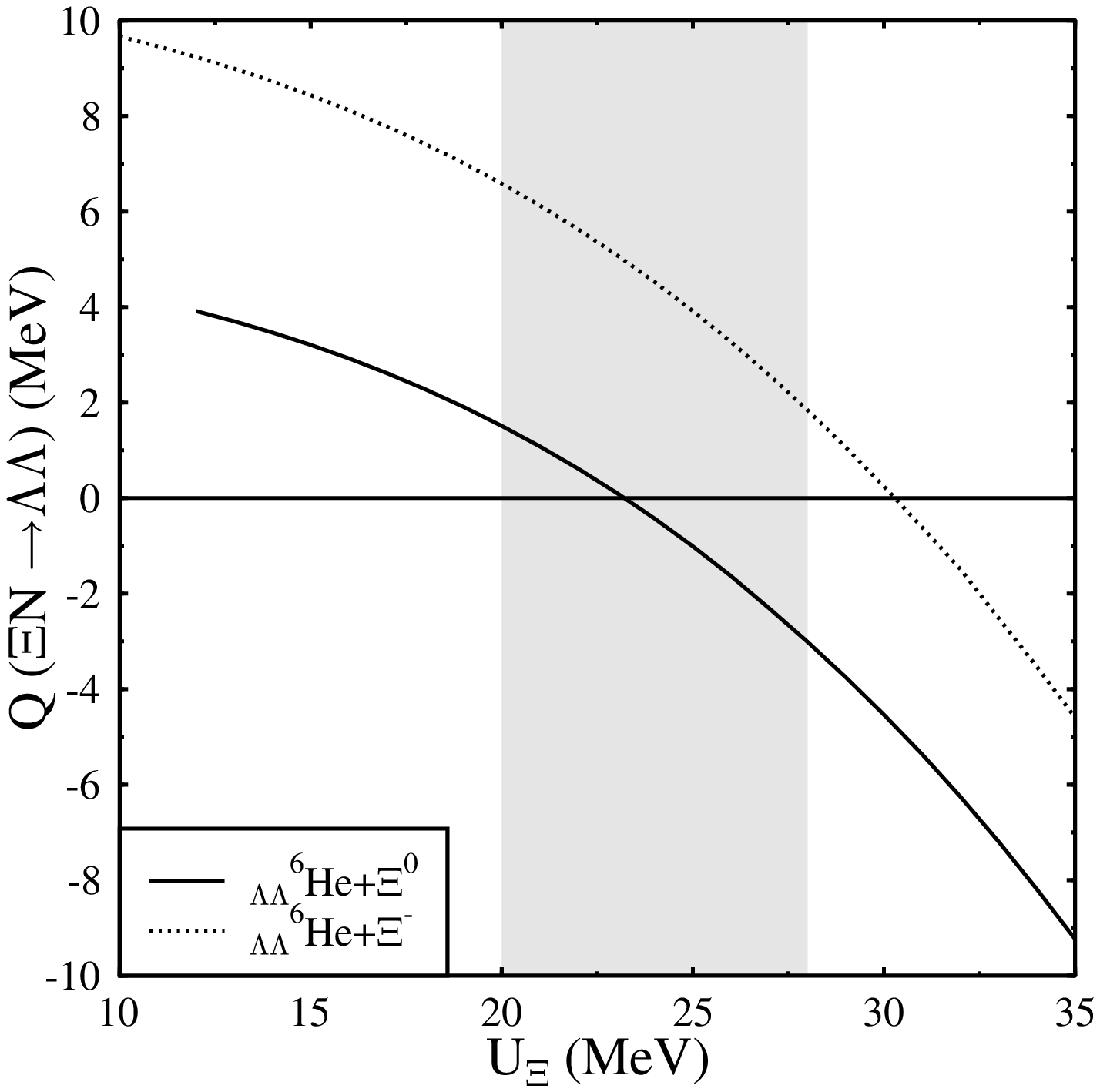,height=2.5in}}%
{The Q-value of the strong process $\Lambda\Lambda\to\Xi$N for
the system with $A=7$ and $S=-4$ as a function of the potential
depth of the $\Xi$ in nuclear matter.}
{pic5}

\subsubsection{Superheavy Hypernuclei}

Now, even heavier systems can be calculated in the extended RMF-model, 
their structure and stability can be studied.
The binding energy of multi-strange nuclei is plotted in
Fig.~\ref{prl2} versus the strangeness fraction $f_s=|S|/A$ for model
1 and model 2 for various nuclear cores.
It is interesting to note that the minima of the curves are located around
$f_s \approx 0.6$ for model 1 and $f_s\approx 1.0$ for model 2, which
is in the same range as for strangelets as discussed below.
A striking example for the enhanced stability is the sequence with a
$^{180}$Th core in model 2. The nuclear core $^{180}$Th consisting of 90 protons and 90 neutrons
is unstable - it is far beyond the proton drip line. 
Nevertheless, the addition of strangeness (hyperons) can stabilize the systems 
and results in binding energies of about $E/A \approx -20$~MeV!\\
\widefig{\psfig{figure=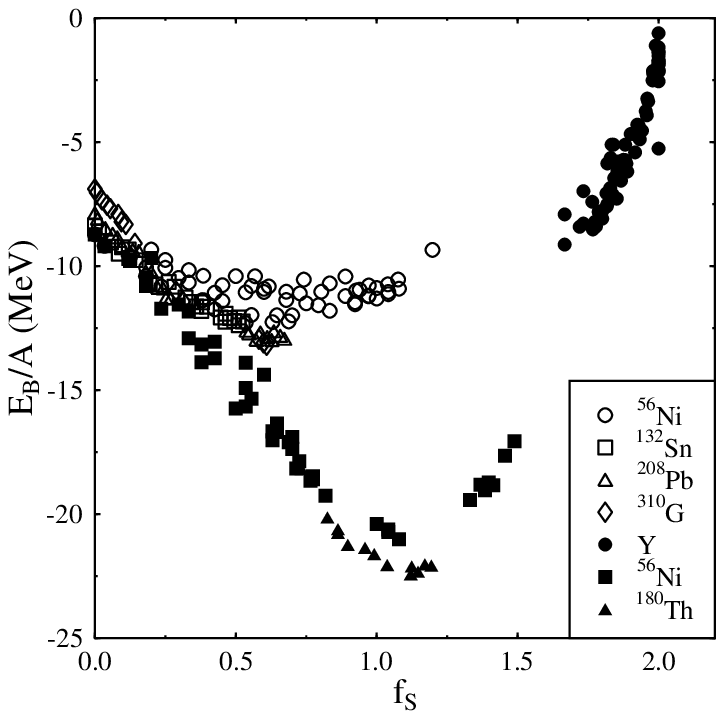,height=2.5in}}%
{Binding energy versus the strangeness fraction for SHM(strange hadronic matter)
in model 1 (nuclear cores $^{56}$Ni, $^{132}$Sn, $^{208}$Pb, $^{310}$G)
and model 2 ($^{56}$Ni,$^{180}$Th, Y denotes purely hyperonic systems).~\cite{Sch93}}
{prl2}

\subsubsection{Pure Hyperclusters - Neutral and Negatively Charged Nuclei}

Figs.~\ref{prl2} and~\ref{prl3} show that in the RMF model exhibits also pure hyperon clusters (denoted as
Y). 
They consist only of $\{ \Lambda,\Xi^0,\Xi^- \}$-hyperons.
Their net strangeness fraction is  $f_s \approx 1.7$ and the charge fraction is $f_q \approx -0.3$.
Pure hyperon clusters are only bound in model 2 due to the 
implemented hyperon-hyperon interaction).
The binding energy per baryon reaches $E/A \approx$ 3 - 8 MeV, $f_s \approx 2$, and $f_q \approx -0.4$.
Fig.~\ref{smallm3} shows the binding energy of MEMOs versus the mass A. 
Note that $E_B/A$ drops linearly with A.
Let us now turn to the dual counterparts of MEMOs, namly the strangelets.\\
\swfig{\psfig{figure=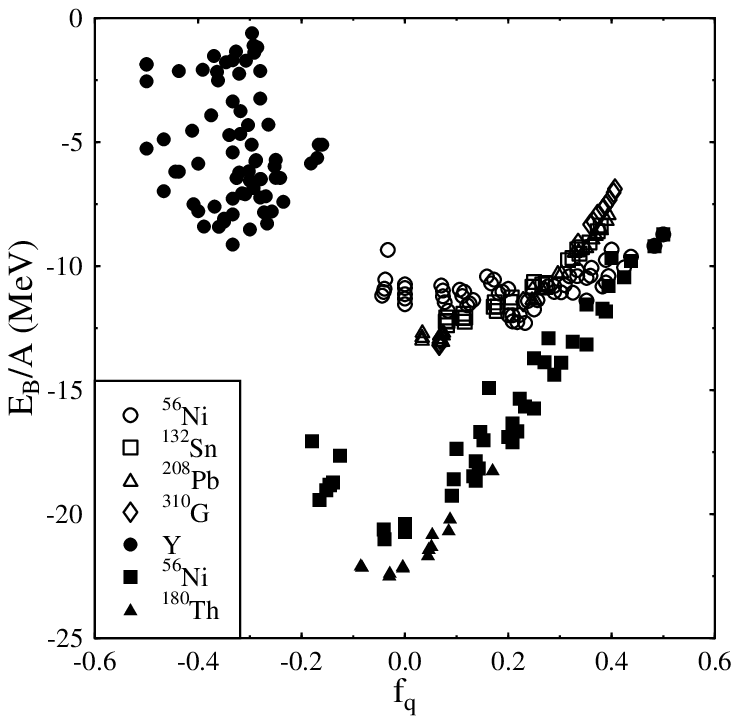,height=2.2in}}%
{5.5cm}
{prl3}{Binding energy per baryon versus the charge fraction $f_q$ for
strange hadronic matter.~\cite{Sch93}}
\hfill
\swfig{\psfig{figure=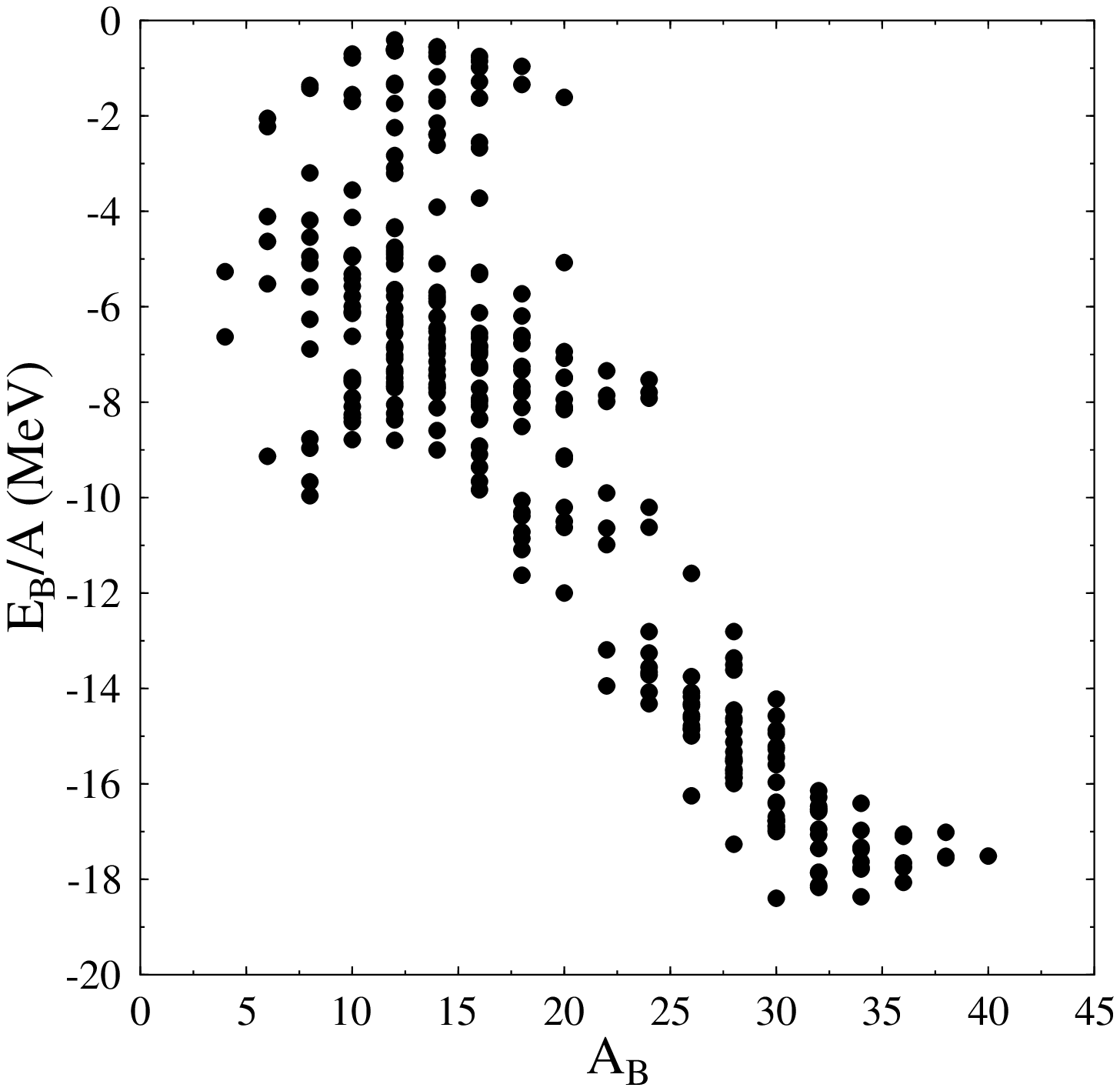,height=2.2in}}%
{5.5cm}{smallm3}{The binding energy per baryon of very light systems of SHM versus the
baryon number $A$.}

\subsection{Strangelets - Strange Quark Matter}

Strange quark matter (SQM) or strangelets are 
strange clusters containing a large number of delocalized
quarks $(u...u\, , \, d...d\, , \, s...s)$, in multiquark droplets.
Multiquark states consisting only of u- and d-quarks have masses considerably
larger than ordinary nuclei.\\
Droplets of SQM, which would contain approximately the same
amount of u-, d- and s-quarks (strangelets, strange multiquark clusters),
might also be denser than ordinary
nuclei. They might exist as long-lived exotic isomers of nuclear matter inside
strange neutron stars.

Speculations on the stability of strangelets are based on the following observations:
\begin{enumerate}
\item The (weak) decay of a s-quark into a
d-quark could be suppressed or forbidden because
the lowest single particle states are occupied
(Pauliblocking, analogous to the MEMO case discussed above). 
\item The strange quark mass can be lower than the Fermi energy of the u-
or d-quark in such a dense quark droplet. Opening a new flavour degree of
freedom therefore tends to lower the Fermi energy and hence also the mass per baryon of the
strangelet (see Fig.~\ref{fig02}), maybe even below the proton mass.
\end{enumerate}
SQM may then appear as a nearly neutral
state (Q$\approx$u+d+s=$\frac{2}{3}e-\frac{1}{3}e-\frac{1}{3}e$=0).\\
\\
\widefig{\psfig{figure=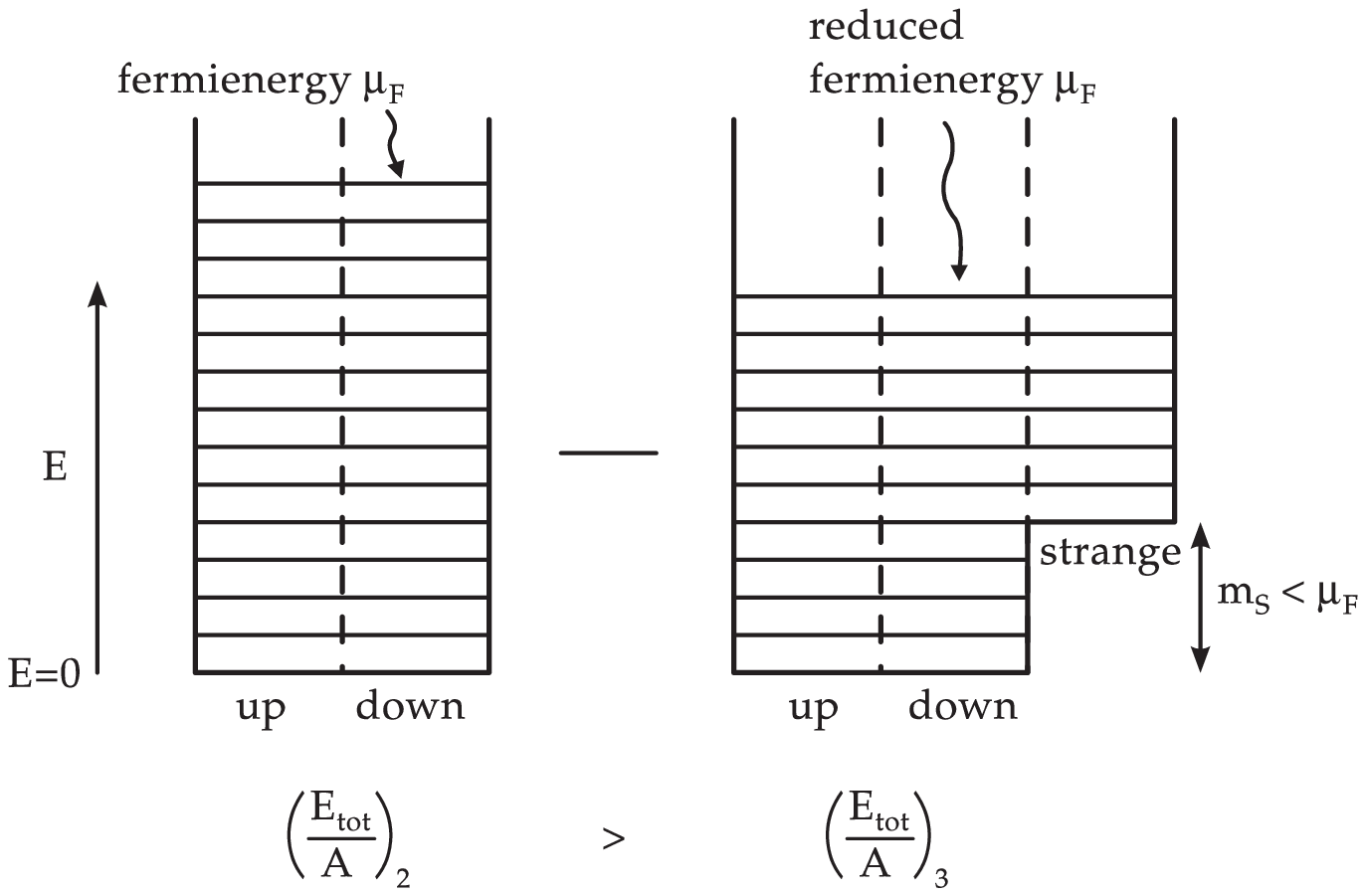,height=2.5in}}%
{Schematic illustration of the energy levels inside a multiquark
 bag with two versus three flavours.}
{fig02}

\subsubsection{Cold Infinite Quark Matter}

Quark matter
in {\em bulk} is often described by an equation of state of interacting quarks to first
order in $\alpha_c = g^2/16\pi $~\cite{Chi79,Bj79}
{\small
\begin{eqnarray}
\label{cg1}
&&\Omega_i(m_i,\mu_i) \,  = \nonumber \\[1em]
&&-\frac{1}{4\pi^2} \,  \left(
\mu_i (\mu_i^2-m_i^2)^{1/2}(\mu_i^2-\frac{5}{2}m_i^2) +
\frac{3}{2} m_i^4 \ln \frac{\mu_i + (\mu_i^2-m_i^2)^{1/2}}{m_i} \right. \\[0.5em]
&&  \, \left. - \, \frac{8}{\pi} \alpha_c \left[
3\, \left(\mu_i (\mu_i^2-m_i^2)^{1/2} \, - \,
m_i^2 \ln \frac{\mu_i + (\mu_i^2-m_i^2)^{1/2}}{m_i} \, \right) ^2
\,  - 2 \, (\mu_i^2 - m_i^2) \right] \, \right) \, \, \, . \nonumber
\end{eqnarray}}
Here $m_i$ and $\mu_i$ denote the (current) mass and the chemical potential,
respectively, of the quark flavour i=u,d,s.
For the total potential the vacuum excitation energy $BV$
has to be added. It corresponds to the energy difference between the
`false', perturbative vacuum inside the `bag' and the true vacuum on the
outside
\begin{equation}
\label{cg2}
\Omega (\mu_q,\mu_s; m_s; \alpha_c) \, = \, \sum_{i=u,d,s} \,
\Omega_i (m_i,\mu_i)  \, + \, BV \, \, \, .
\end{equation}
From this expression (\ref{cg2})
the energy per baryon in the groundstate can be
readily obtained. The derivative of E/A with respect to the baryon density
is zero if the system is at zero pressure. In Fig.~\ref{ta} the resulting groundstates are shown
for $\alpha_c=0$ and different bag-parameters a function of the strangeness
fraction $f_s$. The energy per baryon of the corresponding hyperonic matter
(taken from the RMF model 1) is also drawn. The tangent construction
shown proves that for a given strangeness fraction the separation of a normal and a highly strange subsystem
might be energetically favorable as compared to the mixture. Therefore it is energetically advantageous to accumulate 
the strangeness in the quark phase!\\
\widefig{\psfig{figure=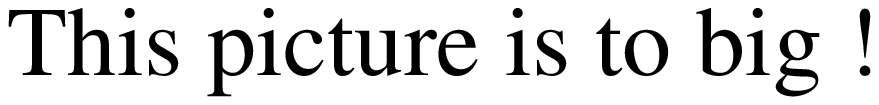,height=2.7in}}%
{The tangent construction at $T=0$.   
For a given strangeness fraction $f_s$
the specific energy $E/A$ is lowest for a two-phase mixture of quark and 
hadronic matter. Drawn are the masses of non-interacting quark matter for
different bag parameters and $m_s=150$~MeV (QGP) and those of infinite
hadronic matter in the RMFT model 1 (full line) and model 2 (dashed)~\cite{diener}.}
{ta}

\subsubsection{Binding Energy per Baryon}

The stability of strangelets depends (in the bag model) strongly on the model 
parameters B and $\alpha_{c}$, but also on the baryon number A. Fig.~\ref{bag}
shows the binding energy per baryon as a function of baryon number.
Note that this model predicts even with the most optimistic parameter set
absolutly stable strangelets only for A$>25$.
 All light strangelets (A$<25$)
are found to be metastable, i.e. they will decay, possibly with typical
weak interaction decay times, $\tau \sim 10^{-10}s$. \\
\widefig{\psfig{figure=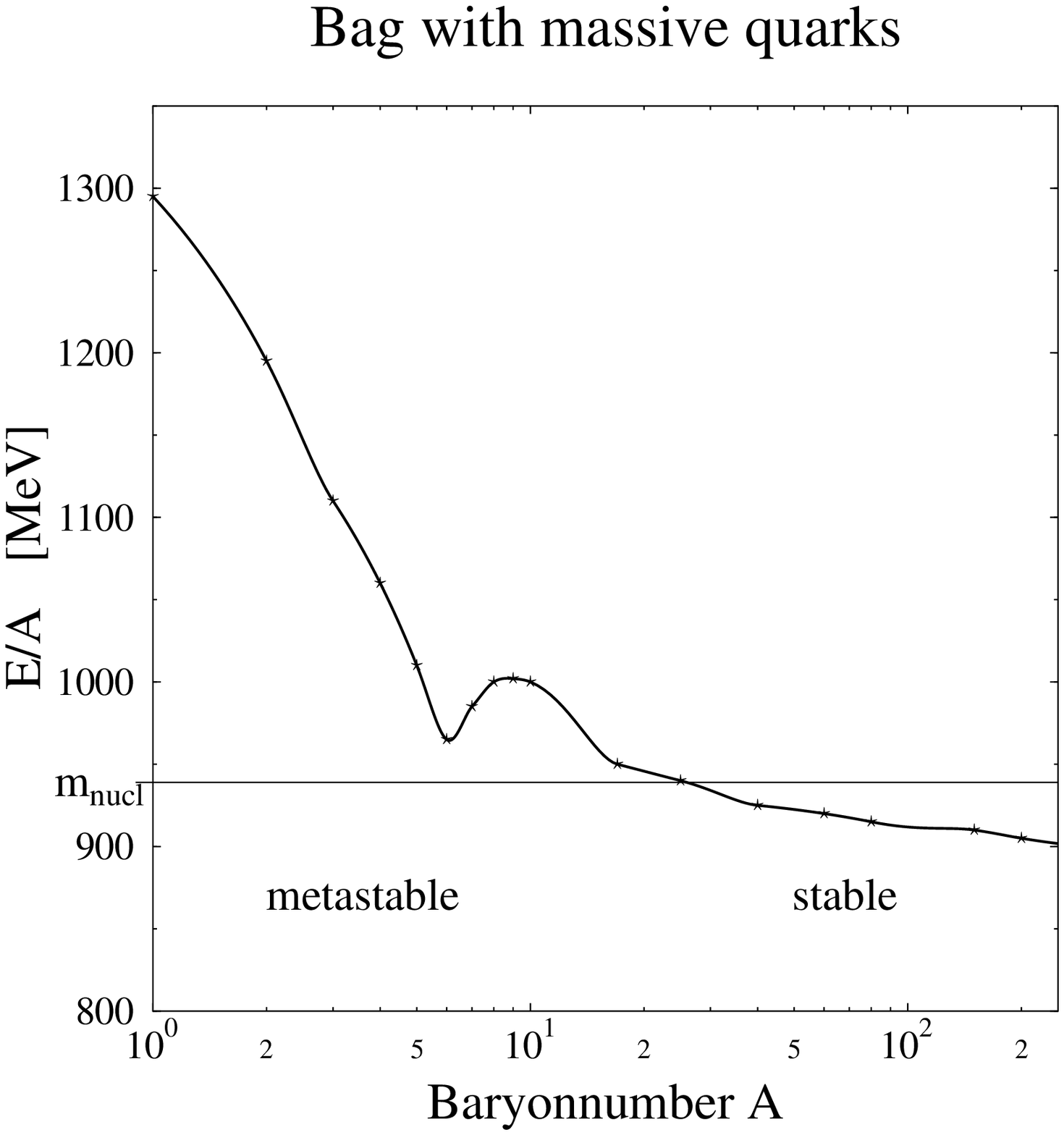,height=2.5in}}%
{Binding energy per baryon as a function of baryon number. The curve has
a local minimum at A$_B$=6, bag constant B$^{1/4}$=145 MeV, m$_s$=150 MeV
and f$_s$=1.0~\cite{PRD91,papa}.}
{bag}
Experiments geared to 
proof the (non)existence of strangelets therefore should clearly cover such
short lifetimes. Unfortunately, this is hard due to the large background of charged hadrons
at the target in violent events with high multiplicities.
To date all experiments concentrate on long flight-path (to minimize background)
and large masses~\cite{pretzl}, although our prediction is that only metastable strange
clusters with $\sim$ cm flight path seem to have a chance of being created.
Coalescence calculations of clusters show that high masses
are suppressed with a penalty factor of $\exp(-A(m-\mu)/T)$,
hence the formation of (even unstable), $A>20$ hyperclusters seems to be highly unlikely.

\subsubsection{Equations of State of Infinite Strangelets}

Let us examine now the phase diagram of a hadronic resonance gas in equilibrium
with the strange quark matter equation of state derived above.
Fig.~\ref{phas} shows the $\rho_{\rm B}-T$ plane for a
fixed strangeness fraction of $f_s=0.1$ and $B^{1/4}=180$~MeV.
Even for low temperatures
and baryon densities $\rho_{\rm B}\approx \rho_0$ this simple model predicts
the coexistence of a hadron- and a QGP-phase.\\
\widefig{\psfig{figure=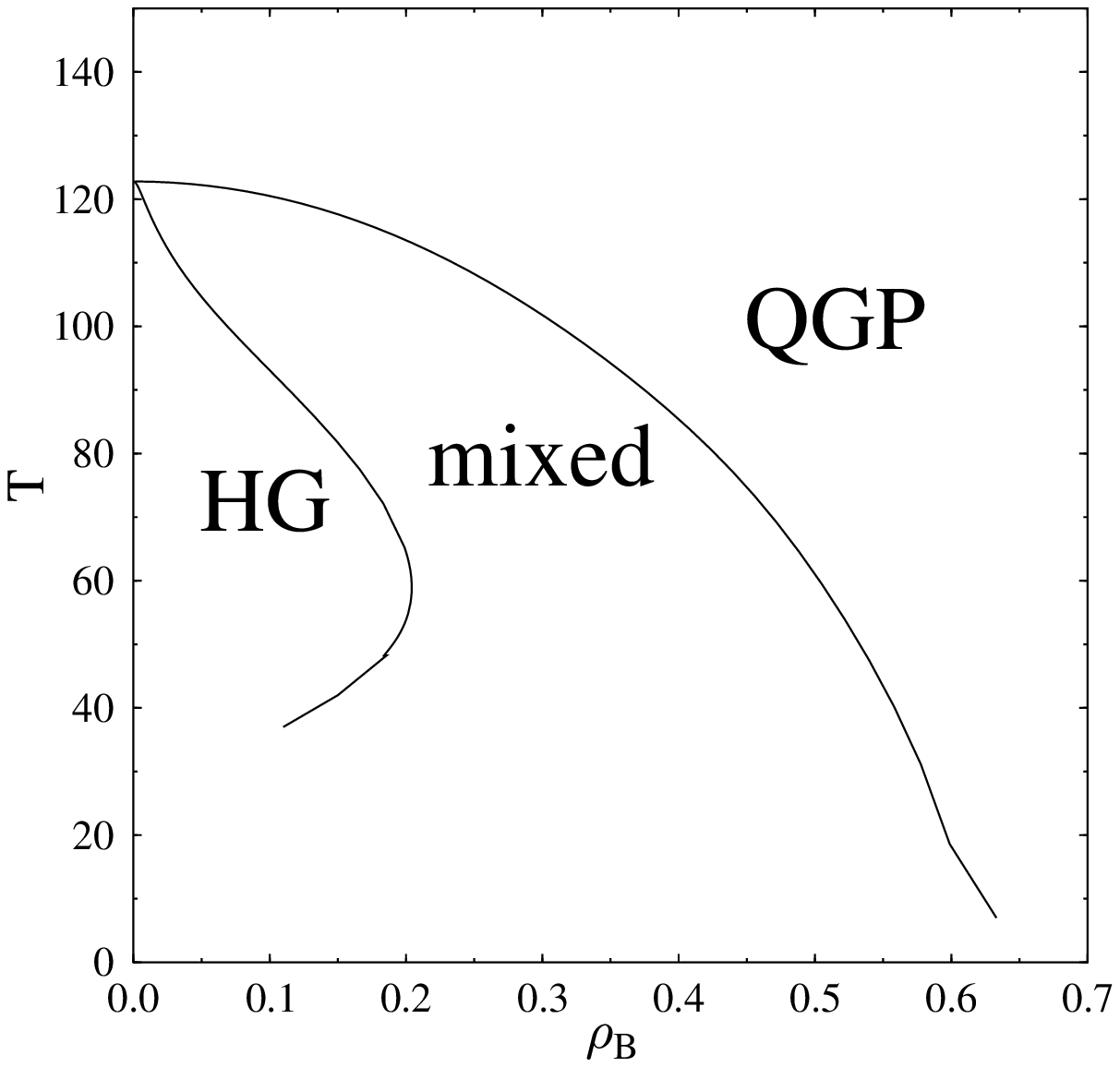,height=2.5in}}%
{Phase diagram of a hadron gas and infinite quark matter (B$^{1/4}=180$MeV)
for fixed $f_s=0.1$.}
{phas}
This illustrates that the thermodynamical properties of a
strange system can indeed be strange! Calculations of QGP signals should therefore take 
this finite net-strangeness degree of freedom into account.

\section{Production of Strange Matter in Heavy Ion Collisions}

After discussing the possible structure and stability of hyperonclusters and 
strangelets, let us turn now to possible mechanisms for their creation.
Leaving aside (for the moment) astrophysical processes (early universe, supernova, 
neutron stars), only relativistic, central heavy ion collisions offer events with sufficiently
large strangeness production (up to several hundred $s \bar{s}$ pairs per event)
to search for multi $s$ (and multi $\bar{s}$!) clusters. Since the distribution in the phase space of
strange and nonstrange particles, is decisive for predictions of hypercluster-probabilities,
let us first study the general single particle distributions in central reactions.

\subsection{Baryon Dynamics in Heavy Ion Collisions}

\subsubsection{Dynamics in a Microscopic Model}

Fig.~\ref{ger1} exhibits the baryon rapidity distribution as predicted by various 
models for heavy ion collisions. ATTILA~\cite{attila} 
and FRITIOF~1.7~\cite{frit} (not in the picture) predict no stopping at SPS(CERN).
They predict
nearly a baryon-free midrapidity already at SPS. This is ruled out by the new CERN data,
which rather support predictions based on the RQMD model~\cite{sorge}. Fig.~\ref{hardy} shows the scaled
final rapidity distribution $\frac{dN}{d(Y/Y_P)}$ (where $Y_P$ is the initial mean rapidity of the projectile)
for the system Pb+Pb, $b<3$fm, at 200 and 1000 MeV calculated with QMD~\cite{aich} and the corresponding reaction at 160 GeV 
calculated with RQMD (histogram).
And also FRITIOF~7.02 version~\cite{fritiof} yields stopping at SPS! Furthermore, even in
very central collisions of lead on lead at $s^{1/2}=6.5$~TeV, there might be some
net-baryon density at midrapidity. This is shown in Fig.~\ref{ger2},
where the event-averaged rapidity densities of net-baryons, hyperons and
anti-hyperons are depicted for the LHC, using FRITIOF~7.02. Even gross features of the 
baryon dynamics are not yet understood. 
\\
\widefig{\psfig{figure=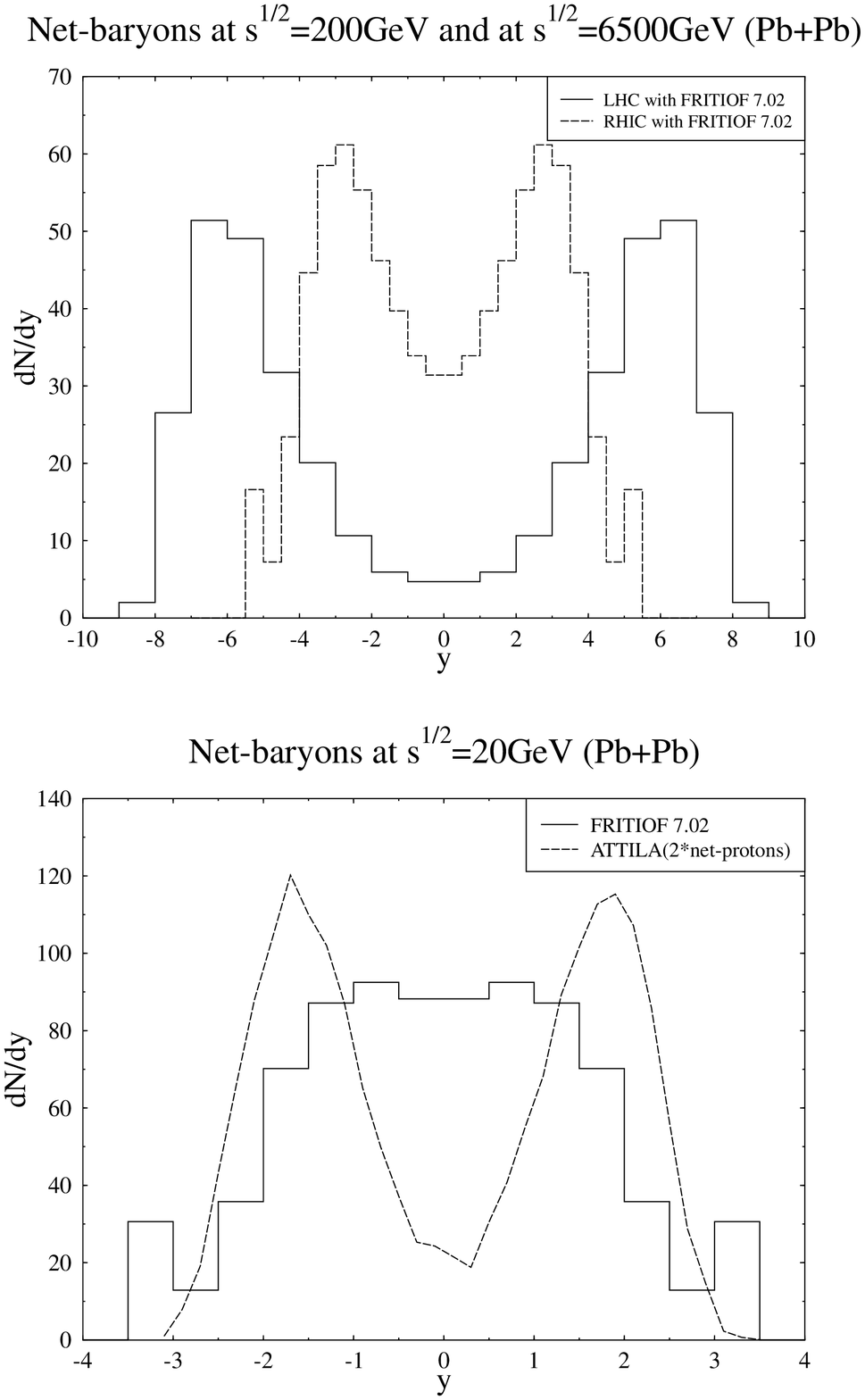,height=3.7in}}%
{Net-baryon rapidity distribution of
very central Pb + Pb collisions at SPS, RHIC, LHC
 calculated with FRITIOF 7.02.
 The midrapidity region
 is even at LHC not baryon-free.}
{ger1}
\widefig{\psfig{figure=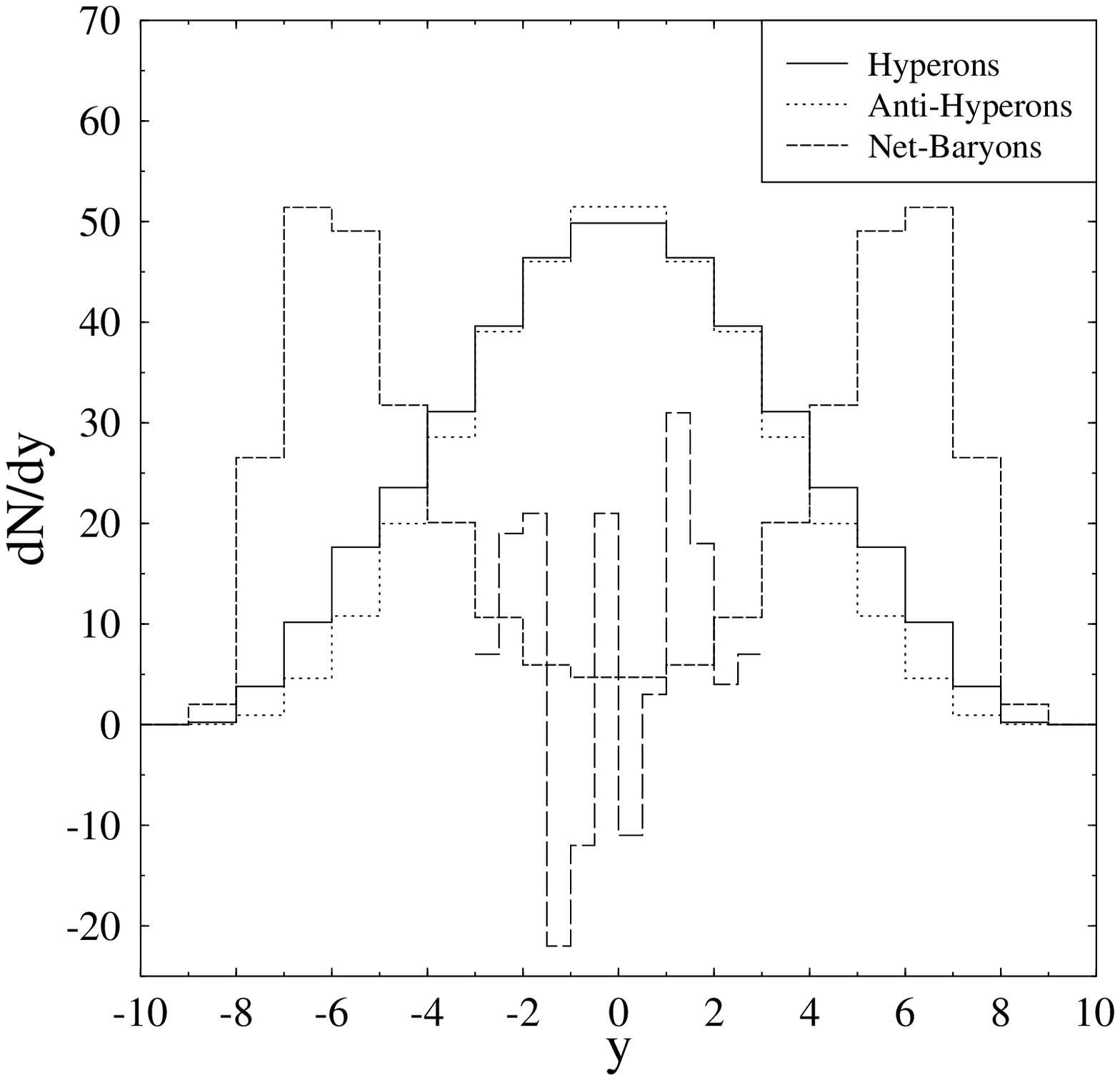,height=2.4in}}%
{The (anti-)hyperon rapidity distribution of very
central Pb + Pb collisions at $s^{1/2}=6500$ AGeV, and
mean net-baryon distribution at midrapidity as compared
with a single event calculated with FRITIOF 7.02.}
{ger2}
\widefig{\psfig{figure=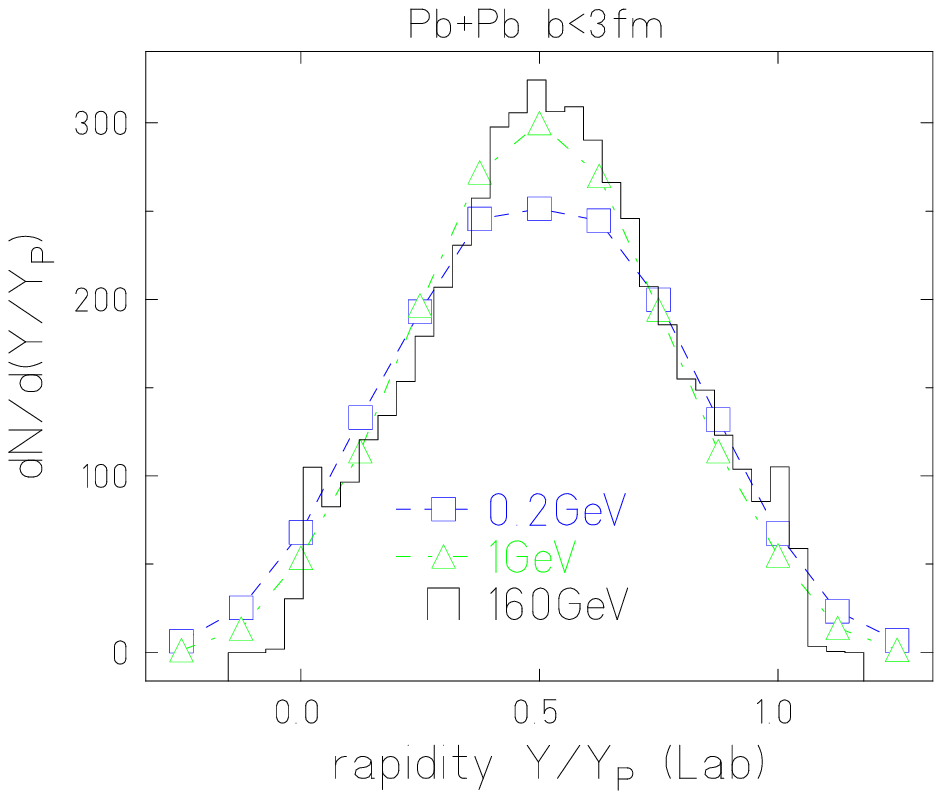,height=2.5in}}%
{Scaled rapidity distribution for the system Pb+Pb $b<3$fm for 200 MeV, 1 GeV and 160 GeV 
incident energy.~\cite{hart}}
{hardy}

\subsubsection{Local Fluctuations}

\widefig{\psfig{figure=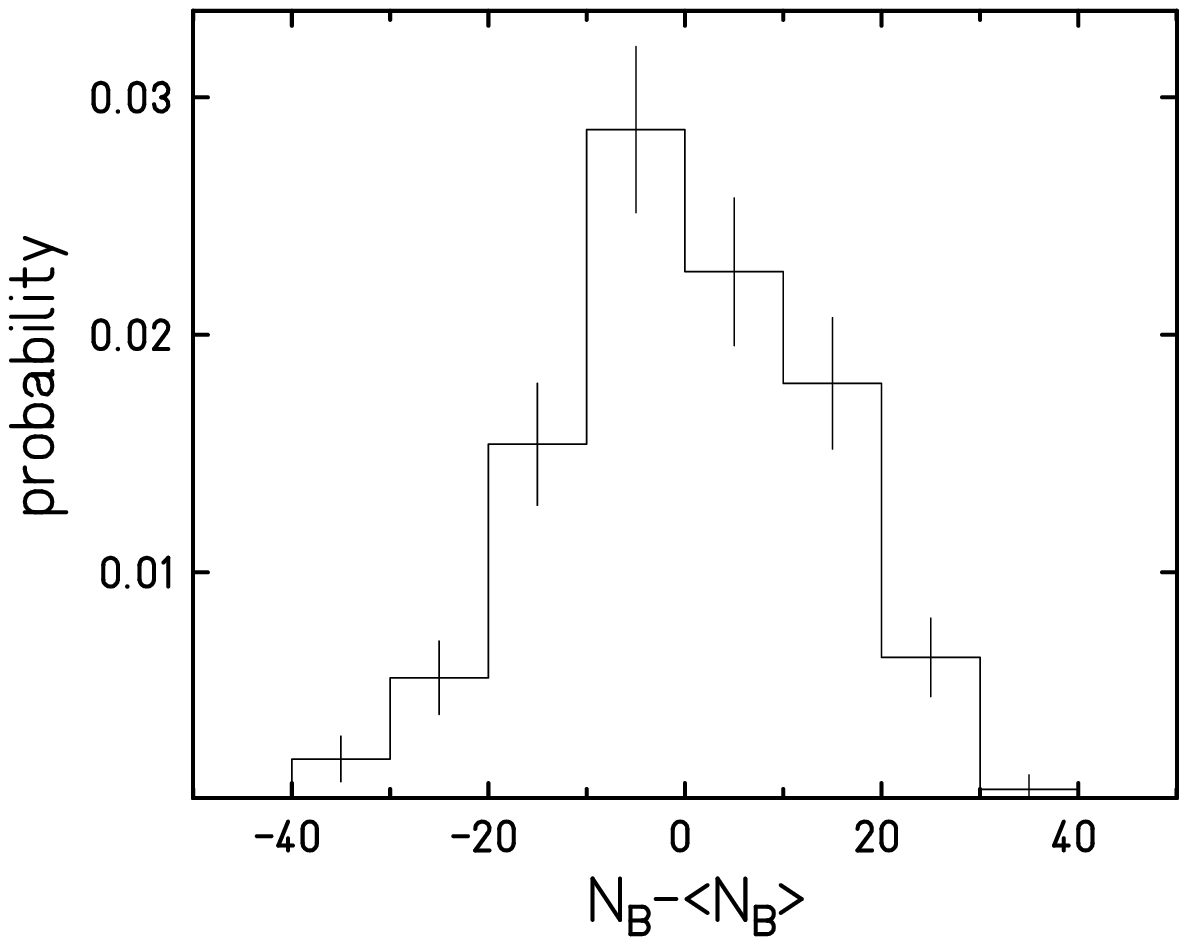,height=2.5in}}%
{Probability distribution for net-baryon number fluctuations at mid-rapidity within bins of one unit
of rapidity width, calculated with FRITIOF 7.02 for the Pb + Pb system at $s^{1/2} = 6500$ AGeV}
{ger3}

To estimate the fluctuations possible in heavy-ion collisions at the LHC,
let us consider a recent Pb+Pb calculation performed at $s^{1/2}=6500$AGeV
with FRITIOF~7.02. The probability distribution for non zero net-baryon number 
fluctuations at midrapidity is plotted in Fig.~\ref{ger3} within bins of one unit
 of rapidity width. This probability is defined for single events, and shows
rapidity density deviations from the average value $<dN/dy>$. 
Bins between $-3<y<3$ have been taken into account. The probability
for fluctuations $N_B-<N_B>$ being larger than $\pm 20$ is about 15 \%. The asymmetry of the 
histogram results from the fact that fluctuations may be different for positive 
and negative deviations around a non zero average rapidity density. 
This aspect is visible clearly in the $dN/dy$ of a randomly   
selected single event as drawn (for the mid-rapidity region) in
Fig.~\ref{ger2}.\\
Keep in mind that the microscopic models used here ignore possible
effects that could change significantly the number of produced strange particles 
in heavy ion collisions, e.g. the string-string-interactions to be discussed now.

\subsection{Medium Effects}

\subsubsection{String-String Interaction}

Fig.~\ref{ger4} depicts the multiplicities of different produced particles at LHC as function of
the string-tension $\kappa$.\\ 
\widefig{\psfig{figure=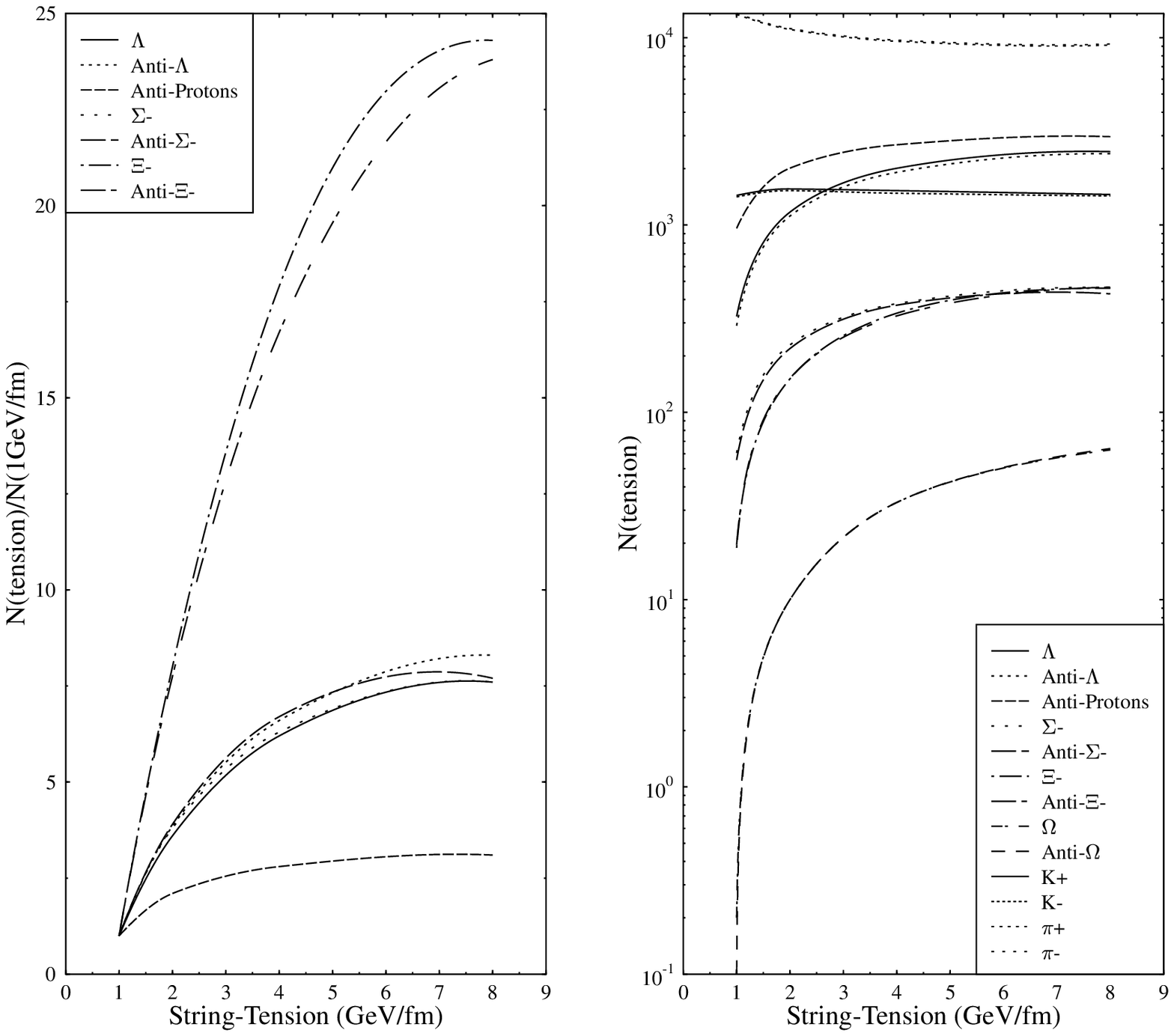,height=5in}}%
{The multiplicities of different particles in very
central Pb + Pb collisions at 6500 AGeV as
function of the string-tension calculated with FRITIOF 7.02.}
{ger4}
The number of produced baryons increases with $\kappa$ while the number of mesons decreases.
This can be read off the formula:

\begin{equation}
\left| M\right|^{2} \propto {\rm exp}\left( -\frac{\pi m^2}{\kappa}\right)\quad.
\label{schwing}
\end{equation}
Here $|M|^2$ is the probability to produce a parton-antiparton pair with the mass m for a
string-tension $\kappa$. This formula is motivated by the Schwinger-formalism \cite{schwinger}
for
pair-production in an infinite electric field.
The string-tension of 1 GeV/fm (no interaction) leads to a suppression of the heavier strange quarks (s)
and diquarks (di), as compared to up (u) and down (d) quarks. The following input is used in our calculations :\\
\\$\frac{|M|_u^2}{|M|^2_u}$ : $\frac{|M|_d^2}{|M|_u^2}$ : $\frac{|M|_s^2}{|M|_u^2}$ :
$\frac{|M|_{di}^2}{|M|_u^2}$ = 1 : 1 : 0.3 : 0.1,\\
\\correspondig to $m_s$ = 280 GeV.
\\A higher string-tension, e.g. 2 GeV/fm yields  
the suppression factors :\\
\\$\frac{|M|_u^2}{|M|^2_u}$ : $\frac{|M|_d^2}{|M|_u^2}$ : $\frac{|M|_s^2}{|M|_u^2}$ :
$\frac{|M|_{di}^2}{|M|_u^2}$ = 1 : 1 : 0.55 : 0.32\\
\\Such enhanced string tensions may effectively simulate
string-string interacions.
\\Another source for the enhanced production of heavy quarks in such reactions can be a reduced
effective quark mass. When a string fragments in a dense medium,
an effective constituent quark mass of say, $m = 0.5m_0$ results in the same suppression factor as a
fourfold increase of the string-tension,
$\kappa = 4$ GeV/fm. The constituent quark masses could be
possibly reduced in the vicinity of a chiral phase transition.

\subsubsection{Reduction of the constituent quark masses}

The relativistic
meson-field models, which, at high temperature,
qualitatitively simulate chiral behaviour of   
the nuclear matter (see next paragraph), exhibit a transition into a
phase of massless baryons \cite{The83}.
Including hyperons and $YY$-interaction \cite{Sch93}
at $\mu\approx 0$,
the densities for all (anti)-baryon species considered are of the order of $\rho_0$ near
the critical temperature.~\cite{spieles}
Thus, the fraction of \mbox{(anti-)}strange quarks
increases drastically.
Several hundred \mbox{(anti-)}baryons, most of them
\mbox{(anti-)}hyperons, may then fill the hot
 midrapidity  region
(with net baryon number $\approx 0$).
Fig.~\ref{allhyp} shows this phase transition for (anti--)nucleons and
(anti--)hyperons at small $\mu$. Below $T_{\rm C} \approx 170$MeV the anti--baryons are
strongly suppressed due to the positive baryochemical potential. The
(anti--)hyperons
suffer an additional suppression because of their higher mass. Above $T_{\rm
C}$ all effective masses are small and the relative yields in the medium
are dictated
by the isospin degeneracy, thus favouring (anti--)hyperonic matter.\\
\widefig{\psfig{figure=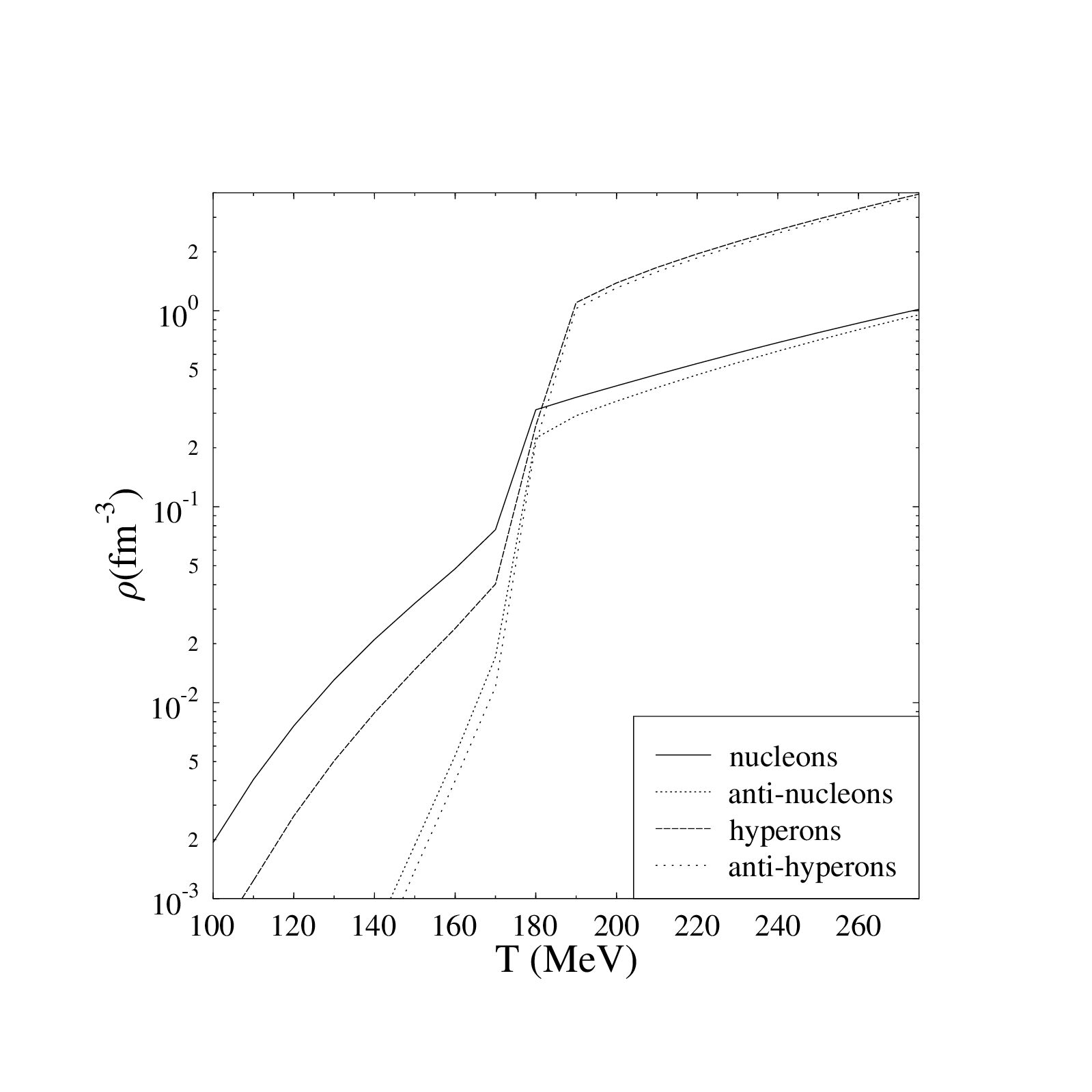,height=2.5in}}%
{Densities of (anti--)nucleons and (anti--)hyperons as functions of
temperature for fixed $\mu_{\rm q}=100$~MeV and
strangeness fraction $f_{\rm s}=0$,
calculated with the relativistic meson--baryon field theory
(RMF model 2).}
{allhyp}
Similar results can be found with a chiral SU(3)$\times$SU(3) Lagrangean~\cite{papa2}.\\
Now that we have demonstrated that a wealth of in-medium effects can increase
drastically the single particle (strange)baryon density, let us consider the clustering of the 
hyperons by using a phase space coalescence model.

\subsection{Clustering of Hypermatter}

The generalized phase space coalescence model has been applied successfully to light (p,n)clusters and anticlusters.
Fig.~\ref{x1x20} shows a calculation of antideuterons~\cite{bleicher}
which exhibit a characteristic phase space structure due to strong absorption.
However, the strange quarks or the hyperons should be  more localized in phase space.
Thinking in terms of a coalescence picture for producing exotic
multistrange clusters, such an early increase in the net strangeness
content, whether a QGP has been formed or not, should affect the production
probability of these clusters.\\
\widefig{\psfig{figure=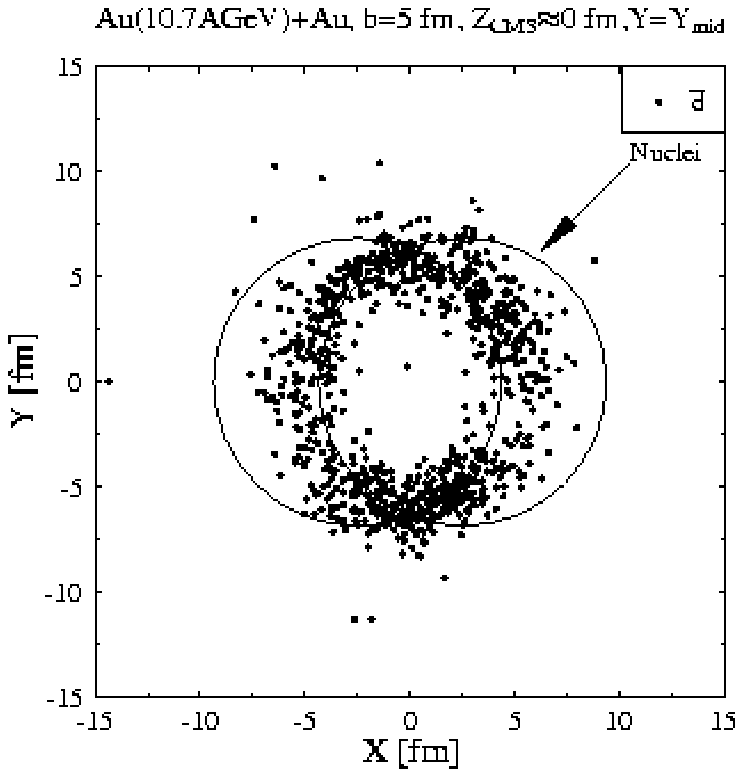,height=2.5in}}%
{Antimatter is only emitted from the surface
of the hot and dense fireball, created in heavy ion collisions~\cite{bleicher}.}
{x1x20}

Microscopic models (Venus, RQMD, FRITIOF,...) do not include the production of light
exotic nuclei dynamically. Cluster formation can be added after the strong freeze-out
i.e. after the last strong interaction of the particle.
The cluster formation probability can be calculated by
projecting the
baryon pair phasespace on the cluster wave function via
the Wigner-function method as described in \cite{rma94,gyu83,ear75}.
The yield of exotic two-particle-clusters is given by
\[ {\rm d}N_{\rm Cluster}={\rm const}  
\Big<\sum\limits_{i,j}\rho^{^{\rm W}}_{\rm Cluster}
(\Delta \vec{R},\Delta \vec{P})\Big>
{\rm d}^3(p_{i_{B,Y}}+p_{j_{B,Y}})\,. \]
The Wigner-transformed wavefunction of the
cluster is denoted by $\rho^{^{\rm W}}_{\rm Cluster}$.
The sum goes over all baryons and hyperons,
whose relative distance ($\Delta \vec{R}$) and relative momentum
($\Delta \vec{P}$) are calculated in their rest frame at the  
time after both nucleons have ceased to interact. The constant
accounts for the statistical spin and
isospin projection on the desired state.
The calculation of higher mass fragments is straight forward by exchanging the two 
particle cluster wavefunction with a n-body harmonic oscillator wavefunction, 
summing up over all possible baryon combinations.
RQMD predicts three $\Lambda \Lambda$- and one $\Sigma^- \Sigma^-$-cluster
in 100 Pb+Pb collisions at SPS.    
An additional strangeness distillation occurs when there is a first order 
phase transition from a quark matter state to a hadron fluid.

\subsection{Strangelet Distillation}

Consider a hadronizing QGP-droplet with net-strangeness zero surrounded by a layer of hadron gas which 
continously evaporates hadrons (they undergo the freeze-out).
Assume the two phases to be in perfect mechanical, chemical and thermal equilibrium.
Now, rapid kaon emission leads to a
finite {\em net} strangeness of the expanding
system \cite{PRD91}.This scenario is visualized in Figs.~\ref{qgp} and~\ref{qgp2}(left).
  This results in
an enhancement of the $s$-quark abundance in the quark phase.
Prompt kaon (and, of course, also pion) emission cools
the quark phase, which then condenses into metastable or stable
droplets of SQM.
\\
\widefig{\psfig{figure=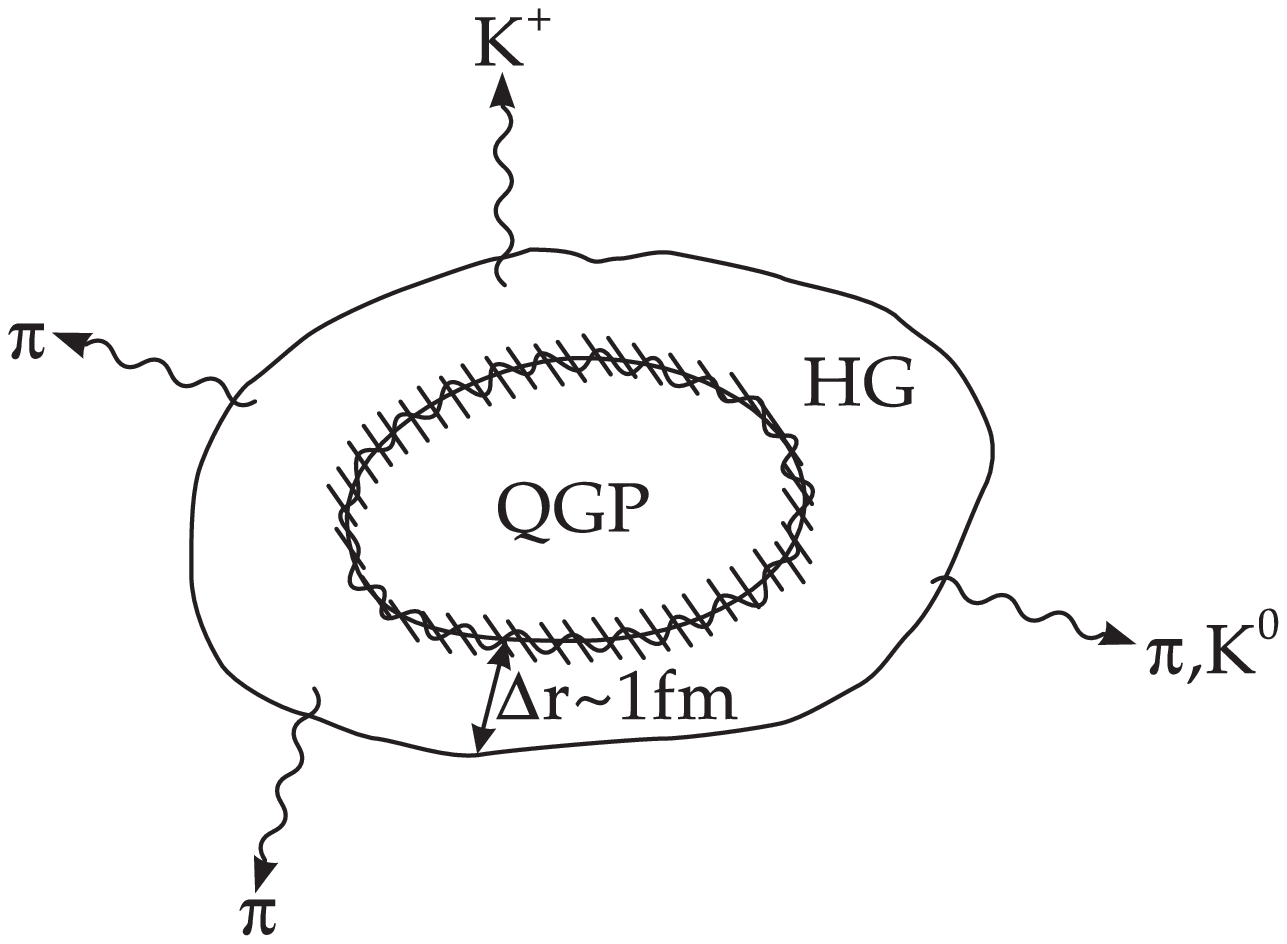,height=2.5in}}%
{Hadron gas surrounds the QGP at the phase transition.
Particles evaporate from the hadronic region.
New hadrons emerge out of the plasma by hadronization.}
{qgp}

\subsubsection{Distillation at High Baryon Densities}

Fig.~\ref{qgp2} (right) shows the properties of a strangelet in its time evolution
for two different bag constants and a moderate initial specific entropy of
$S/A=25$.
 The ratio of the
quarkchemical potential and the temperature $|\mu | /T$ is directly related
to the entropy per baryon number via
\begin{equation}
\left(\frac{S}{|A_B|}\right)^{\rm QGP} \, \approx \, \frac{37}{15} \pi^2 \,
\left(\frac{|\mu |}{T}\right)^{-1}
\,\,\, .
\end{equation}

Therefore, we deal with rather high net baryon densities in the system.
Strangelets can survive if they are allowed to cool down. In any scenario,
strangeness is accumulated in the quark phase.
From this it seems misleading to extract "temperatures" and "chemical potentials"
from detected particle ratios, since their abundances reflect time integrated values,
not a specific break-up $\mu/T$ combination.
 The rates of the emitted hadrons are dictated by the (strange) chemical
potentials which are strongly time dependent.\\
\\
\widefig{\psfig{figure=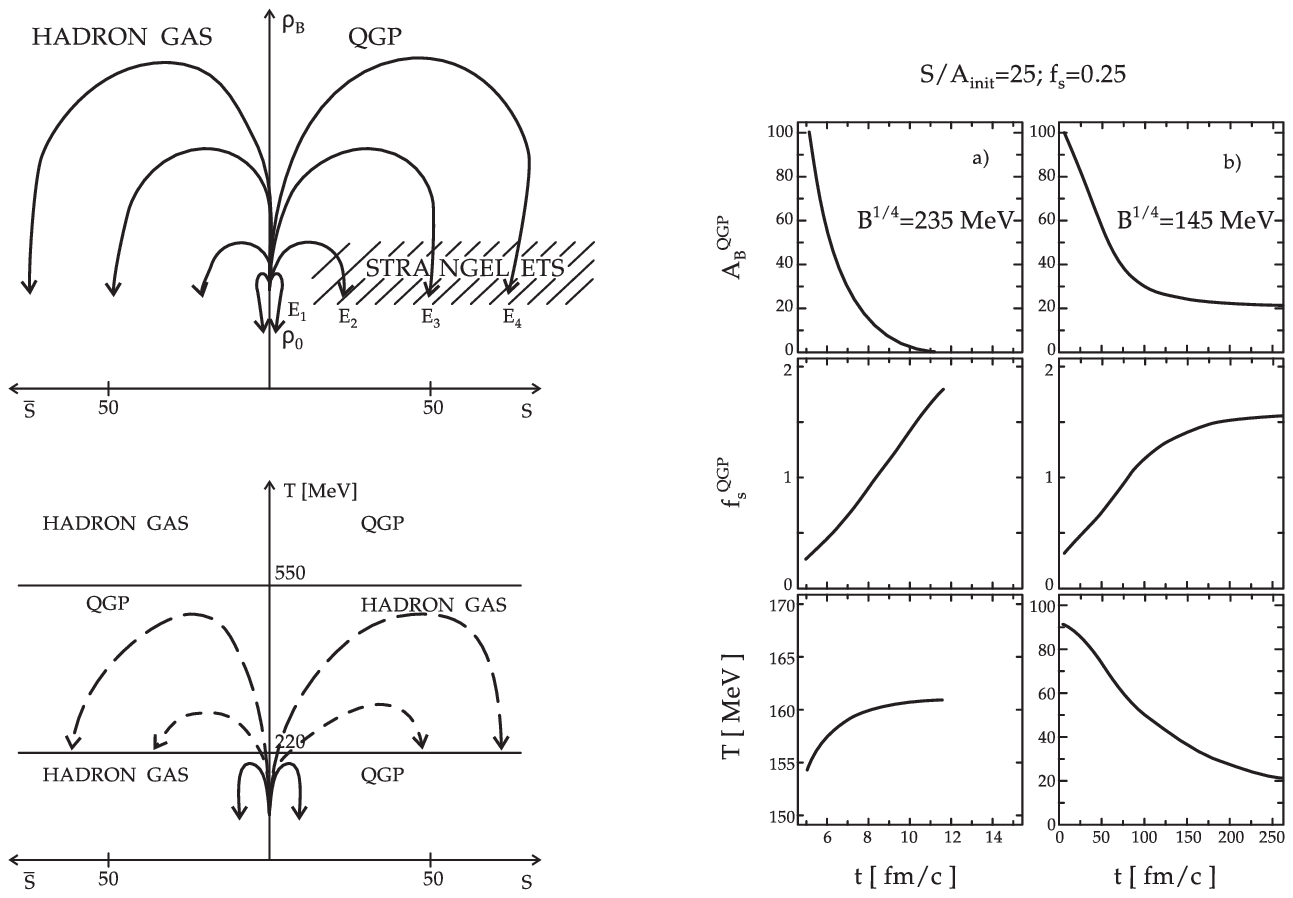,height=3.4in}}%
{Left : The fountain of strangeness production~\cite{diener}. The distillation of strangeness 
is more effective for high baryon densities.
Right~\cite{PRD91} : a) Baryon number, strangeness content and temperature of the
quark glob during complete hadronization as a function of time
for a very large bag constant
$B^{1/4}=235$ MeV.
The initial values are an initial baryon content of $A_B(t_0)=100$,
an entropy per baryon ratio of
$S/A(t_0)=25$ and an initial net
strangeness fraction of
$f_s(t_0)=0.25 $.
Note the strong increase of the strangeness content with time.
\newline
b) The same situation as in a), however, for a small bag constant
$B^{1/4}=145 $ MeV, when a strangelet is distilled.
One observes a strong decrease in the evolving temperature.}{qgp2}
Fig.~\ref{fig14} illustrates that the different species are radiated off after different 
times: e.~g. the lambdas stem from the very last stage while the
kaons are emitted very early.\\
\widefig{\psfig{figure=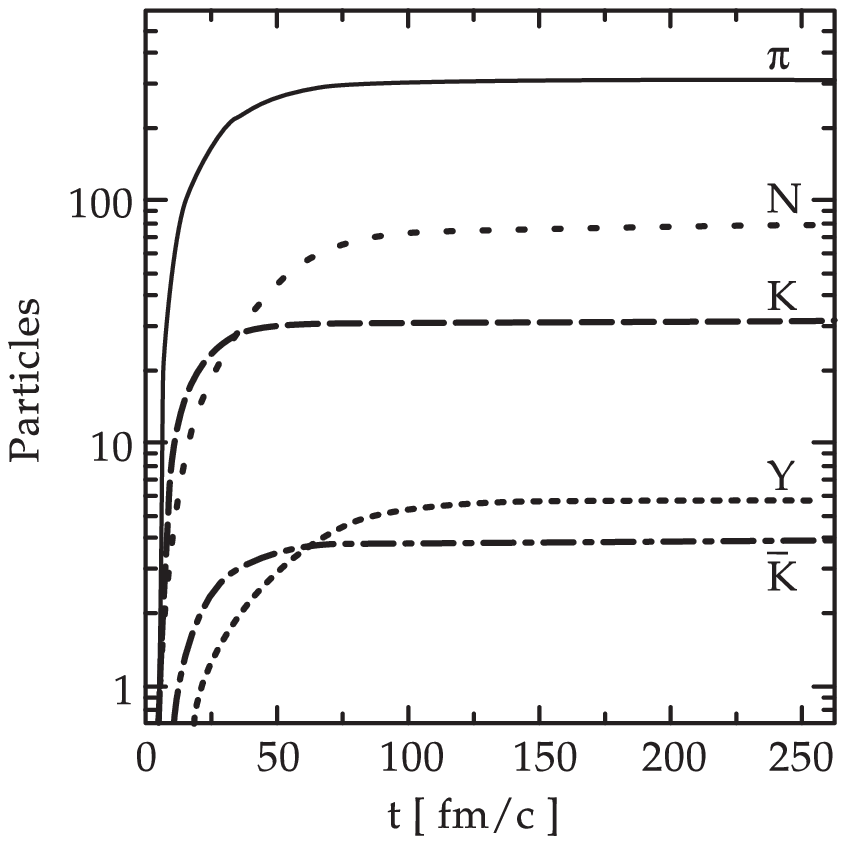,height=2.5in}}%
{The number of emitted particles are shown versus time in a situation,
when a strangelet is distilled. The conditions are the same as in the
previous figure b), however, no initial net strangeness is assumed.~\cite{PRD91}}
{fig14}

\subsubsection{Distillation at low $\mu/T$}

At collider energies, the midrapidity region is expected to be characterized 
by low net-baryon densities. How can one expect to create stable strangelets
with baryon density of  $\rho > \rho_0$? 

Fig.~\ref{tdens} illustrates
the increase of baryon density in the plasma droplet
as an inherent feature of the dynamics of the phase
transition.
This result originates from
 the fact that the baryon number in the
quark-gluon phase is carried by quarks with $m_{\rm q}\ll T_{\rm C}$, while
the baryon density in the hadron phase is suppressed by a Boltzmann factor
$\exp (-m_{\rm baryon}/T_{\rm C})$ ($m_{\rm baryon}\gg T_{\rm C}$).
A very tiny excess of initial net-baryon number will suffice to generate regions of
very high density $\rho_{\rm B}>\rho_0$! The very low initial $\mu/T$
corresponds to high values of the initial specific entropy.

Fig.~\ref{fsrho} shows the evolution of the two-phase system for $S/A^{\rm
init}=200$, $f_s^{\rm init}=0$ and for a bag constant $B^{1/4}=160$~MeV 
in the plane of the strangeness fraction vs. the baryon density.
The baryon density increases by more than one order of magnitude!
Correspondingly, the chemical potential rises as drastically
during the evolution, namely from $\mu^i=16$~MeV to $\mu^f>200$~MeV.\\ 
\vspace{-1cm}
\widefig{\psfig{figure=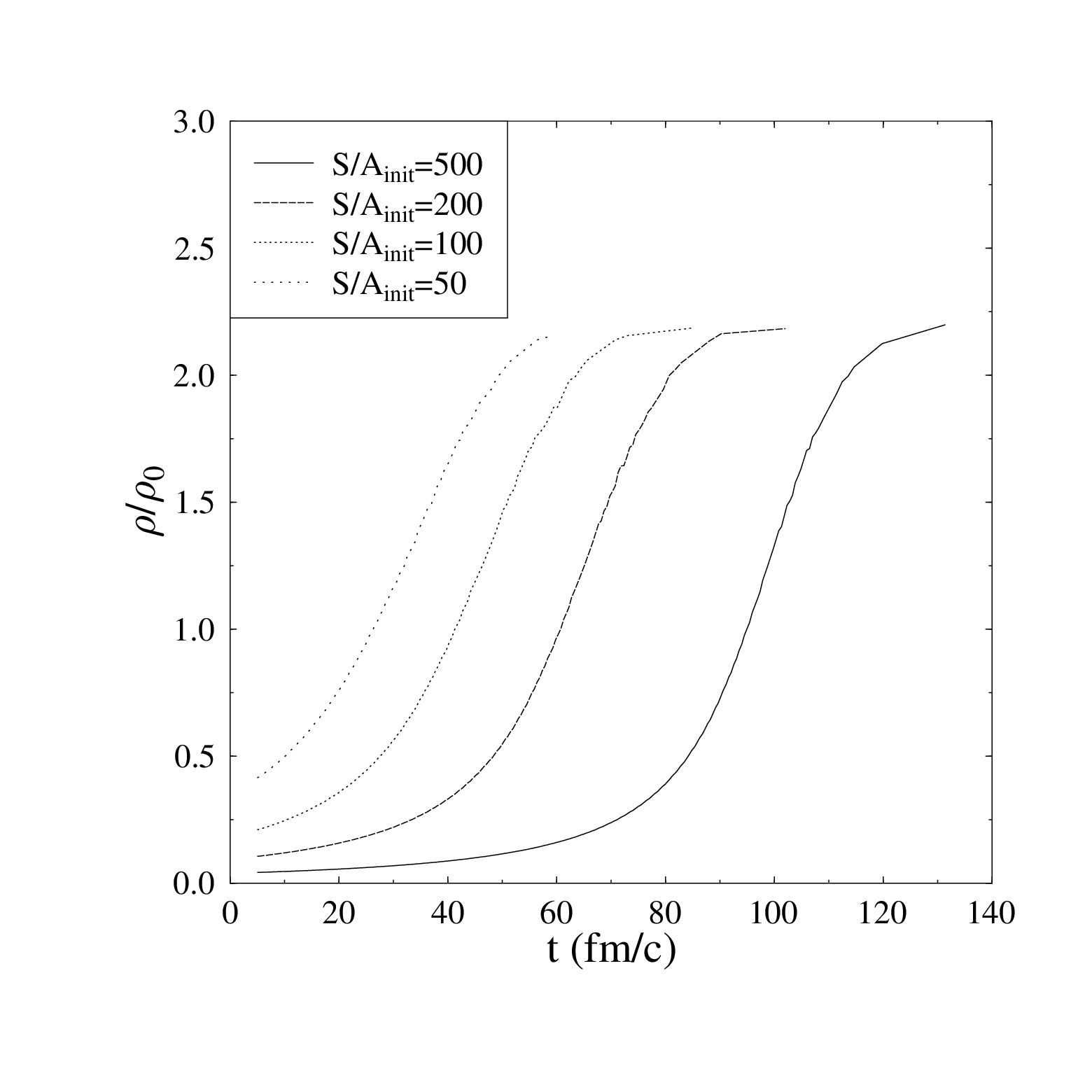,height=3in}}%
{Time evolution of the net baryon density of a QGP droplet.
The initial conditions are
$f_s^{\rm init}=0$ and $A_{\rm B}^{\rm
init}=30$. The bag constant is $B^{1/4}=160$~MeV.~\cite{spieles}}{tdens}
\hspace{-8mm}   
\widefig{\psfig{figure=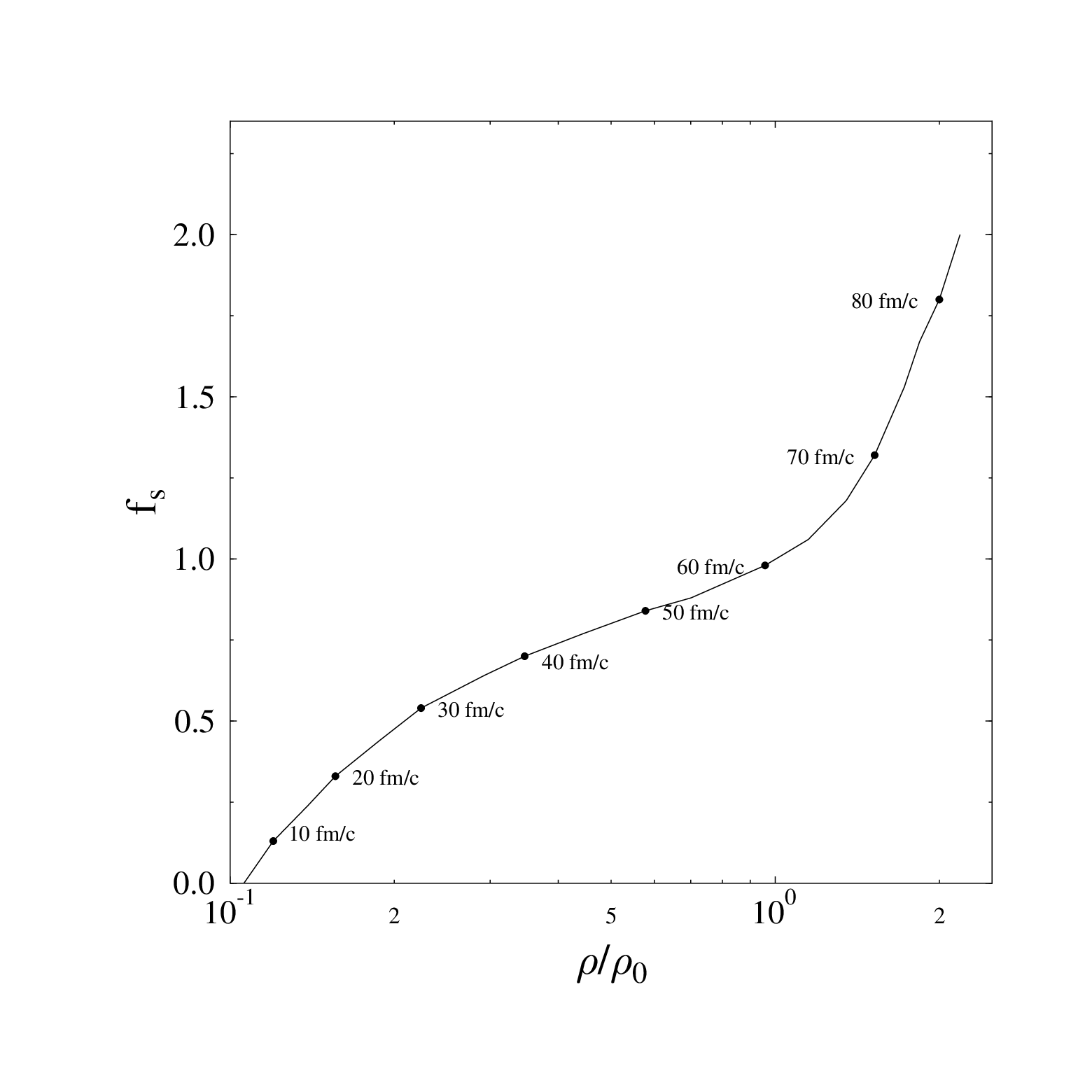,height=3in}}%
{Evolution of a QGP droplet with baryon number $A_{\rm B}^{\rm
init}=30$ for $S/A^{\rm init}=200$ and
$f_s^{\rm init}=0$. The bag constant is $B^{1/4}=160$~MeV. Shown is the
baryon density and the corresponding strangeness fraction.~\cite{spieles}}{fsrho}
The strangeness
separation mechanism~\cite{cg} drives the chemical potential of the strange quarks
from $\mu^i_s=0$ up to $\mu^f_s\approx 400$~MeV.
Thus, the thermodynamical and
chemical properties during the time evolution
differ drastically from
 the initial values! 
Low initial chemical potentials do not hinder the creation
of strangelets with high $\mu$.

\section{Experiments - On the Search of Strange Matter}

\subsection{Present experiments -- the search for absolutely stable strange matter}

The key for the detection of strange (quark) matter is the predicted low  charge-to-mass ratio.
In contrast to normal nuclei with $Z/A=0.5$, strange matter should have charge-to-mass ratios as 
small as $\pm1:10$ or $\pm1:20$.\\
Strangelets and MEMOs in flight 
are practically indistinguishable from each other experimentally, but they can be unambigously 
separated from normal (anti-)matter.\\
Searches for strange quark matter in Au+Au
collisions at the AGS have been peformed~\cite{SA}, but only for a small acceptance close to zero degrees, for  
charge-to-mass ratios higher than 1:25.\\
Large acceptance data will be gathered at Brookhaven with the E864 detector~\cite{sand}.\\
At CERN a search for strange matter has been started using a 500 m long setup. 
First lead on lead data have yielded upper limits of $10^{-7}$ per event for strange matter.

The sensivity of the experiments can be increased by several orders of magnitude. However, all these
experiments are unable to observe metastable hyperclusters due to the required lifetimes $\tau \gg 10^{-10}$s.
This is dictated by the long flight path in these experimentes. 

\subsection{Future experiment -- searching for shortlived strange matter}

The lifetime of a MEMO or a strangelet could be similar to the $\Lambda$ 
lifetime ($\sim 10^{-10}$ s). Thus, a short flight path, open geometry
detector will be needed to discover these objects. If a
produced strangelet is absolutely stable with respect to strong interaction, the only energetically possible
decay mode is the weak leptonic decay ($s \rightarrow d$, $Q \rightarrow
 Q'+ e + \bar{\nu }$), which will transmute the strangelet to the strangelet minimum
energy. The time-scale for this weak process
has been estimated to be $\sim  10^{-4}$ s~\cite{Chi79}.\\
But in case that strangelets or MEMOs are metastable (this is the case in a much wider range of parameters)
the weak conversion rate $u+s \leftrightarrow u+d$ for
almost cold SQM can be calculated \cite{Ko92}. It is $\sim 10^{-6} -  
10^{-7}$ s.
 Subsequent weak    
decay processes, as described here, `heat' up the droplet (on a scale of
a few MeV). The droplet may be cooled by $\gamma $-radiation or by
nucleon emission. Both processes compete with each other. Still, the
lifetime issue is not settled.

\section{Conclusion}

Strange-baryon yields can go up  by orders of magnitude at $T_c$ due to changes in the string tension or
 the effective mass at a phase transition (e.g.in chiral models).
The clustering probability for strange matter should then increase correspondingly.

For strong decays, the short lifetime could not allow to see strangelets and MEMOs
directly. \\
However, measurements of Hanburry-Brown-Twiss-correlations between
two or more hyperons can be applied to prove the (non)existence of strange matter.
To achieve sensitivity in detectors for two (or more) strange particles per event, the singles
rate must be high, though.

Negatively charged hyperclusters (e.g. $Z=-2,f_s=2$) might exist with positive baryon number.
Their experimental signature in magnets should be similar to that of anti-nuclei($\overline{\rm He}$),
however, they would {\bf not} show the additional annihilation energy of antinuclei
in calorimeters. 

Baryochemical potentials may turn out to be nonzero (even at $y_{cm}$) at colliders, with $\mu/T \sim 1$.
The event by event rapidity dependence of $\pi$,p,$\bar{\rm p}$,Y and $\overline{Y}$ should be measured to clearify this issue.
\\This could also reveal fluctuations of the (strange) net-baryon number.

Even the small, but finite stopping power of baryons at LHC can provide phase space regions with $\mu>T$.

The baryon distillery mechanism drives the chemical potential from $\mu_B^i,\mu_s^i \\ \approx 0$
to $\mu_B^{Qf} \gg \mu_B^i $, $\mu_S^{Qf} \sim \mu_B^{Qf}$. If such a first order phase transition exists, the quark phase charges up rapidly with strangeness. 
The system will therefore not stay in the $\rho$,T($f_s$=0) plane, but will follow a complicated three-dimensional path in the space ($\rho$,T,$f_s$). This can
affect severely many of the QGP signals discussed to date. Unfortunately almost nobody has yet undertaken such a study.
Finite $f_s$-values change the phase structure completely! The coexistence region then ranges possibly down to $\rho_B \approx \rho_0$ and $T \approx 20$MeV.

 The bag term is crucial for the time evolution of the QGP.\\ 
For $B^{1/4} \ge 180$ MeV the lifetime of the mixed phase can be large! Multi-$\Lambda$-evaporation
 at the end of the hadronization will signal  by HBT- (or multi- correlation-)analysis the decaying short-lived "strangelet".
For $B^{1/4} < 180 $ MeV metastable strangelets ( Z $\approx$ -2 ?) can be formed and might be detected directly.


\end{document}